# The economy-wide rebound effect and U.S. business cycles: A time-varying exercise[*]

Marcio Santetti[†]

## Abstract

Energy efficiency gains in production and consumption are undisputed economic and environmental goals. However, potential energy savings derived from efficiency innovations may have short-lasting effects due to increased demand for more affordable energy services. Measuring the size of this rebound effect is a critical tool for better assessing the reliability of energy-saving technological change for global warming mitigation. This paper estimates the size of the economy-wide rebound effect using time-varying Vector Autoregressive (VAR) models with stochastic volatility for U.S. business-cycle peak and trough periods. All models estimate a rebound effect close to 100%, with reductions in energy use lasting no longer than three years following energy efficiency innovations. The latter, therefore, are an insufficient tool for effectively changing historical energy use patterns.

---

[*]I would like to thank Adalmir Marquetti for providing insightful comments and suggestions. The usual disclaimer applies.

[†]Emerson College. Email: marcio.santetti@emerson.edu.

# 1   Introduction

An effective mitigation of global warming implies that economies must decrease their greenhouse gas emission rates while still achieving positive national income growth. This absolute decoupling requires energy efficiency improvements that increase the ratio of final outputs relative to the demanded energy inputs. Energy-saving technological change, therefore, is expected to lead the transition to a more sustainable growth paradigm from economic and institutional standpoints (Brockway et al., 2021; U.S. International Energy Agency, 2025).

Improved energy efficiency, however, may have the opposite effect. The expected positive impacts of newer technologies may either be partially or completely offset by a rising demand for energy services. The notion of a "rebound effect" suggests that the response of economic agents—consumers, producers, and governments—to lower effective energy prices resulting from energy-saving innovations can erode the latter's potential contributions for reducing emissions (Gillingham et al., 2016).

This work explores the macroeconomic dimensions of the rebound effect. I estimate the size of the economy-wide rebound effect for the U.S. economy using Vector Autoregressive (VAR) models with time-varying coefficients and stochastic volatility. In a model with energy use, economic activity, and real energy prices, I compute the rebound effect as the response of energy use to a one-standard deviation exogenous, unanticipated shock to energy efficiency that is allowed to change over time.

Recents works have applied VAR methodologies to estimate the economy-wide rebound effect via impulse-response analysis (Bruns et al., 2021; Berner et al., 2022). One limitation of standard VAR approaches is that these compute unique parameters describing the dynamic interactions between two or more time series that supposedly represent an entire sample period. Thus, these do not take into account, for example, how parameter values and volatilities may endogenously change according to different cycles. The works of Primiceri (2005) and Del Negro and Primiceri (2015), primarily employed in empirical macroeconomics, provide efficient estimation algorithms based on Bayesian methods which allow for a time-varying behavior in the VAR system's data-generating process. To the best of my knowledge, this is the first work using this methodology to compute rebound effects.

Results and conclusions can be summarized as follows. All estimations suggest that almost the entirety of energy savings are put back into the economic system via increased energy use approximately three years after energy efficiency shocks. Models in monthly frequency for business-cycle peaks and troughs show that the largest economy-wide rebound effects happened in the 1990s, with net increases in energy demand. Quarterly estimates, although relatively smaller in magnitude and variance, suggest a similar pattern. Overall, this paper's time-varying economy-wide rebound effects are smaller than those estimated by the other aforementioned VAR applications. However, all reach a similar policy conclusion: energy-saving technological change is by itself an insufficient economic and institutional tool for an effective departure from historical rebound and emission trends.

In addition to this Introduction, this paper has four more sections. Section 2 further explores the rebound effect's concept and surveys the applied literature quantifying its size. Section 3 describes the data set, the time-varying VAR methodology, and the identification strategy utilized to derive energy-saving innovations and the posterior responses of energy use. Section 4 presents and discusses the econometric results. Lastly, Section 5 concludes.

## 2 Related literature

Energy efficiency gains in production and consumption are undisputed economic and environmental goals, but properly assessing its overall repercussions can be paradoxical and difficult to measure. On one hand, businesses, households, and markets benefit from energy-saving innovations that make energy services more affordable. On the other, these savings may trigger opposite reactions: increases in energy demand may either partially or completely offset newly implemented efficiency improvements. This "rebound effect" suggests that behavioral and economic changes in the demand for energy induced by energy-saving innovations may waste the latter's potential benefits.

Academic discussions on the rebound effect date back to the late 1970s. The seminal paper by Khazzoom (1980) introduced the idea that gains from energy efficiency mandates—for cars and household appliances, for example—may be counterbalanced by increases in overall energy use. Since then, the existence of this phenomenon, although conceptually consensual, has been mixed on modeling and empirical grounds (Brookes, 1990; Grubb, 1990).

This section briefly surveys the literature on the macroeconomic dimensions of the rebound effect. Also known as the economy-wide rebound effect, it is the result of direct and indirect rebound events (Barker et al., 2009). Direct rebound effects arise from an increase in the demand for a specific energy service following an efficiency improvement in its own sector (e.g., electricity). With the fall in its effective price, both households and firms tend to consume more of that energy source than prior to the innovation.[1] Indirect rebound effects result from changes in the demand for other goods and services in response to an energy-saving innovation. For example, improvements in transportation from more fuel-efficient vehicle engines may increase consumers' demand for cars, thus ramping up firms' demand for steel, rubber, plastic, and copper, among other inputs, all of which require energy for their provision.

---

[1] Microeconomic treatments of the rebound effect are vast. See, for example, Greening et al. (2000), Sorrell and Dimitropoulos (2008), and Azevedo (2014).

After an energy-saving innovation is put in practice, its actual efficiency may not equal its full potential, since a share of savings may be put back into the system within its specific sector or in the production of other goods, services, and energy carriers. Therefore, the economy-wide rebound effect is measured as the percentage of the expected energy saving that is not effectively realized due to direct and indirect rebound effects (Turner and Hanley, 2011; Bruns et al., 2021). It can be expressed as follows:

$$\text{Rebound effect (\%)} = \left[1 - \frac{\text{Actual \% reduction in energy use}}{\text{Expected \% change in energy use}}\right] \times 100 \qquad (1)$$

From Equation (1), a rebound effect of 100% means that the same amount of energy has been put back into the system following the adoption of a new energy-saving technique. Values lower than 100% indicate an actual reduction in energy consumption after an innovation. Lastly, a rebound greater than 100% implies an actual net increase (backfire) in energy demand after an efficiency improvement.

The most common methodology for quantifying the economy-wide rebound effect are Computable General Equilibrium (CGE) models. These simultaneous-equation systems are based on social accounting matrices, and its parameters express structural features of countries, regions, or sets thereof. As its name suggests, this procedure simulates how an entire system reacts to shocks (disturbances) in specific variables, such as energy productivity/efficiency parameters. In these models, the rebound effect can be calculated by comparing the system's response to improvements in an energy input parameter vis-à-vis a counterfactual model where no innovation happened in that sector (Brockway et al., 2021).

In addition to geographic locations, studies in this applied field vary on whether energy efficiency improvements are made by households or businesses. More specifically, different authors model energy improvements as either increases in energy efficiency adopted by firms or via increased consumers' utility from energy-related innovations. Regardless of focus, the majority of CGE models study the rebound effect by assuming a "one-off" shock to energy productivity, instead of a permanent disturbance.[2]

Allan et al. (2007) use UKENVI, a model specifically parameterized for the United Kingdom comprehending 25 sectors. These include five energy industries: coal, gas, oil, and renewable and non-renewable electricity. The authors simulate a 5% energy efficiency shock to all sectors, estimating a 30% rebound effect in the short and 50% in the long run.[3] In other words, a 5% improvement in energy efficiency leads to a reduction in energy use of 3.5% in the short and of 2.5% in the long-run.

---

[2]An exception to this shock approach is the model in Wei and Liu (2017), who assume annual global improvements in energy productivity.

[3]The authors assume that the long run comprehends about 25 years, over which the capital stock can be updated.

Similar short-run values are found by Anson and Turner (2009). The authors impose a 5% energy efficiency increase shock to the Scottish commercial transport sector. Focusing on innovations specific to the oil industry, short- and long-run economy-wide rebound effects are 36.5% and 38.3%, respectively. Guerra and Sancho (2010) carry out an empirical exercise for the Spanish economy, estimating an 87% long-run economy-wide rebound effect after a 5% efficiency improvement.

Kulmer and Seebauer (2019) model household behavior. The authors set up a general equilibrium model for Austrian households with heterogeneous preferences facing a 10% efficiency increase in fossil-fuel consumption. Their baseline model returns an economy-wide rebound effect of 65%, where direct effects range between 8% and 12%, with the remainder stemming from indirect events. Furthermore, the paper's main model infers that a 43% tax rate on fossil fuel use neutralizes overall rebound effects.

Also focusing on consumers, Duarte et al. (2018) evaluate economic and environmental prospects of a 20% reduction in electricity and vehicle fuel use for Spain. Using data between 2005 and 2015, their three-scenario analysis delivers an average 75% economy-wide rebound effect. Although their results align with Spanish environmental goals for 2030, the authors remark that severe behavioral changes must be adopted to realize these scenarios.

As remarked by the literature survey in Brockway et al. (2021), results produced by CGE models are divergent not only due to different authors' geographical foci or sectoral specifications but also to this methodology's own limitations. For instance, CGE approaches generally assume perfectly competitive markets, utility-maximizing consumers, and constant returns to scale. However, the most relevant source of wide-ranging results concerns the models' calibration and parameter values. Results are usually highly sensitive to assumed functional forms, arbitrary choices of parameter and elasticity values, and model closures.

In addition to CGE models, recent contributions have been based on econometric analysis. Brockway et al. (2017) make use of a Constant Elasticity of Substitution (CES) production function with three inputs: capital, labor, and energy. While the authors nest the first two factors, the last is left unnested. Furthermore, the energy input is replaced by the narrower definition of "exergy" (that is, available energy to perform physical work). Using non-linear techniques and bootstrapping, the authors estimate economy-wide rebound effects of 13% for the U.S. and the U.K. and of over 100% for China between 1980 and 2010. These numbers are derived from the elasticity of exergy use with respect to exergy efficiency parameters.

Using a sample 55 countries for the 1980–2010 period, Adetutu et al. (2016) use stochastic frontier analysis to, first, estimate an energy efficiency parameter and, second, a dynamic panel model to obtain the elasticity of energy demand with respect to energy efficiency. The authors derive a short-run 90% rebound and a negative 36% long-run effect. This latter negative value is an exception in the applied literature (Brockway et al., 2021). Wei et al. (2019) start off from an aggregate Cobb-Douglas production function and find rather inflated economy-wide rebound effects across 40 countries between 1995 and 2009, such as a 300% rebound for South Korea, 232.8% for Turkey, and 104.3% for the United States.

Bruns et al. (2021) and Berner et al. (2022) use Vector Autoregressive (VAR) techniques to empirically estimate the economy-wide rebound effect for the United States and European countries. The first work uses monthly and quarterly U.S. data on primary energy use, real energy prices, and Gross Domestic Product (GDP) as endogenous variables in a Structural VAR (SVAR) setting. Arguing that theoretical premises in the energy economics literature are uncertain, the authors identify their models using Independent Component Analysis (ICA), a machine-learning procedure for finding the most stable matrix configurations to derive structural shocks (innovations). Over the 1973–2016 period, their models estimate a 100% rebound effect after four years.

The second paper proposes an improvement over the research described above. It identifies a Structural Factor-Augmented VAR (S-FAVAR) model addressing possible omitted-variable bias by adding several potential confounding factors to properly estimate the economy-wide rebound effect. Among these, the authors include exchange rates, industrial production indexes, price expectations, unemployment rates, and producer price indexes for several sectors. With France, United Kingdom, Germany, and Italy added along with the U.S. economy to their sample data, the authors' estimates for the economy-wide rebound effect range from 78% to 101%.

Both Bruns et al. (2021) and Berner et al. (2022) calculate the rebound effect in a particular way. Using impulse-response analysis, these works identify an energy efficiency shock as energy use's own-response to an unanticipated, structural shock. Since their models control for other confounders, they claim that this innovation properly reflects improvements in the energy sector. One methodological limitation of these two works lies in the assumption that their estimated parameters describe an entire sample period. Standard VAR models do not take into account potential time variations in the responses of their endogenous variables to different shocks. Since the works of Primiceri (2005) and Del Negro and Primiceri (2015), efficient estimations of VAR models with time-varying parameters have become extensive in the empirical macroeconomics literature. To the best of my knowledge, this is the first attempt at applying their technique in the energy economics literature to quantify economy-wide rebound effects.

In summary, the growing empirical literature on the macroeconomic rebound effect is diverse in its methods, locations of interest, choice of agents, and, most importantly, results. An important takeaway from these works is that estimations embracing a large set of countries tend to inflate rebound effect values compared to single-economy studies. Furthermore, multi-sector/country models must take into account the distinct energy mix configurations for different economies, especially on their relative share of renewable resource use. This paper focuses on a single economy, the United States, directly taking off from the works of Bruns et al. (2021) and Berner et al. (2022) in the application of VAR models to identify energy efficiency innovations. The novelty here, presented in Section 3, lies on the proposed methodology that aims to capture the time-varying behavior of the economy-wide rebound effect for different U.S. business cycles.

## 3 Empirics

This section describes the empirical methodology employed to measure the economy-wide rebound effect for the U.S. economy. It also briefly describes the data and its sources, as well as the proposed identification strategy.

### 3.1 VAR methodology

The rebound effect may be estimated by how energy demand reacts to an exogenous, unanticipated shock to energy efficiency. To that end, Vector Autoregressive (VAR) models are a suitable empirical technique, as not only do these estimate dynamic interactions between two or more variables over time but also allow for the analysis of exogenous shocks to the variable system (Lütkepohl, 2006).

Following Pfaff (2008) and Nakajima (2011), a VAR process of lag order $p$ can be written in structural form as follows:

$$A\mathbf{y}_t = A_1^* \mathbf{y}_{t-1} + \cdots + A_p^* \mathbf{y}_{t-p} + \varepsilon_t \qquad (2)$$

where $\mathbf{y}_t = (y_{1t}, y_{2t}, ..., y_{Kt})'$ is a row vector of $K$ observed endogenous variables; $A$ and $A_i^*$ are $K \times K$ matrices of instantaneous and lagged effects, respectively; and $\varepsilon_t$ is a $K$-dimensional vector of white-noise structural residuals (innovations).

Assuming a lower-triangular form for the $A$ matrix in (2) and left-multiplying both sides by $A^{-1}$, one has:

$$\mathbf{y}_t = A_1 \mathbf{y}_{t-1} + \cdots + A_p \mathbf{y}_{t-p} + \mathbf{u}_t \qquad (3)$$

where $A_i = A^{-1} A_i^*$ for $i = 1, ..., p$, and $\mathbf{u}_t = A^{-1} \varepsilon_t$. Equation (3) is, then, a reduced-form VAR($p$) process.

This methodology serves many purposes such as inferring Granger causality, variance decomposition, forecasting, and impulse-response analyses. The latter, in particular, quantifies how much one of the VAR endogenous variables responds to a one-standard deviation exogenous shock to itself and/or another variable in the system. Impulse-response analysis thus econometrically translates the idea of a possible rebound effect when energy efficiency projects are implemented: it is the response of energy demand to a one-standard deviation shock to energy efficiency (Bruns et al., 2021; Berner et al., 2022).

One limitation of the standard VAR approach from Equations (2) and (3) is that its coefficients, by assumption, do not change over time. In reality, the mean and variance of a time series may change endogenously over a sample period, and the same applies to parameters describing the dynamic interactions between different time series. In this context, statistical models that allow for a time-varying parameter behavior have been proven useful in empirical macroeconomics (Stock and Watson, 2007; Koop et al., 2010; Clark and Ravazzolo, 2015). In the energy economics literature, though, this approach is still scarce.[4]

In the context of VAR methods, time-varying increments first appeared in Cogley and Sargent (2001) and Sims (2001). In an application to U.S. monetary policy, Primiceri (2005) lays out a VAR model whose interaction and variance parameters are allowed to change over time following random walk processes. Its estimation uses an efficient algorithm based on Markov Chain Monte Carlo (MCMC) methods. This algorithm was later corrected by Del Negro and Primiceri (2015) and has since become a widely used procedure.[5]

The reduced-form model from Equation (3) becomes a time-varying VAR($p$) process as follows:

$$\mathbf{y}_t = A_{1,t}\mathbf{y}_{t-1} + \cdots + A_{p,t}\mathbf{y}_{t-p} + \mathbf{u}_t \tag{4}$$

where now the elements of $A_{i,t}$ (denoted by $\beta_t$) are time-varying interaction coefficients. Primiceri (2005) further assumes that the reduced-form shocks contained in $\mathbf{u}_t$ may be heteroskedastic with variance-covariance matrix $\Omega_t$.

This aforementioned assumption of heteroskedasticity is known as stochastic volatility. As such, the first and second moments of variables' responses to exogenous shocks are allowed to change over time. As reminded by Nakajima (2011), stochastic volatility addresses potential biases from using time-varying coefficients. The variance-covariance matrix $\Omega_t$ can be reduced in the following expression:

$$B_t \Omega_t B_t' = \Sigma_t \Sigma_t' \tag{5}$$

where $B_t$ is a triangular matrix, whose non-zero elements are denoted by $\alpha_t$. $\Sigma_t$ is a diagonal matrix, whose log-transformed elements are $\sigma_t$.

---

[4]One example is the work of Ajmi et al. (2015), who study Granger causality relations involving carbon dioxide emissions, energy use, and aggregate income using a time-varying approach for the G7 countries.

[5]See, for example, Galí and Gambetti (2015); Carriero et al. (2018); and Ali et al. (2023).

Finally, A VAR($p$) process with time-varying parameters and stochastic volatility can be written, in reduced form, as

$$\mathbf{y}_t = A_{1,t}\mathbf{y}_{t-1} + \cdots + A_{p,t}\mathbf{y}_{t-p} + B_t^{-1}\Sigma_t \mathbf{u}_t, \tag{6}$$

with all parameters in the $A$, $B$, and $\Sigma$ matrices following first-order random walk processes:

$$\boldsymbol{\beta}_{t+1} = \boldsymbol{\beta}_t + u_{\boldsymbol{\beta}_t} \tag{7}$$
$$\boldsymbol{\alpha}_{t+1} = \boldsymbol{\alpha}_t + u_{\boldsymbol{\alpha}_t} \tag{8}$$
$$\boldsymbol{\sigma}_{t+1} = \boldsymbol{\sigma}_t + u_{\boldsymbol{\sigma}_t}, \tag{9}$$

where $\boldsymbol{\beta}_t$, $\boldsymbol{\alpha}_t$, and $\boldsymbol{\sigma}_t$ are jointly independent and identically normally distributed terms.

### 3.2 Data

I estimate VAR models with time-varying parameters and stochastic volatility across different periods in the U.S. economy using monthly and quarterly data. Following applied works such as Adetutu et al. (2016) and Bruns et al. (2021), all models' identification strategy reflects the dynamic interactions involving three variables: energy use ($e$), real economic activity ($y$), and real energy prices ($pr$).

In line with the applied energy economics literature, I use the following time series for the upcoming VAR estimations:

- Energy consumption, $e$: Monthly total primary energy consumption (measured in quadrillion BTU), retrieved from the U.S. Energy Information Administration (EIA), Table 1.1. The series was seasonally adjusted using the X-13-ARIMA-SEATS method (U.S. Census Bureau, 2023);

- Real activity, $y$: Monthly and quarterly GDP data derived from the Brave-Butters-Kelley Leading index (measured in standard deviation units from trend real GDP growth). This cyclical measure was initially published by the Federal Reserve Bank of Chicago and is currently maintained by the Indiana Business Research Center at the Kelley School of Business, Indiana University;

- Real energy prices, $pr$: Monthly U.S. crude oil composite acquisition cost by refiners (measured in dollars per barrel), retrieved from the U.S. EIA. Nominal prices were adjusted for inflation using the U.S. Consumer Price Index (CPI), with December 2024 as the base period.

For robustness purposes, I use two additional measures of real activity:

- Monthly index of global real economic activity (measured in percent deviations from trend), proposed in Kilian (2009) and corrected in Kilian (2019). This alternative economic activity index is based on percentage changes in ocean shipping freight rates for commodities such as coal, fertilizer, grains, and oilseeds (Kilian and Zhou, 2018). For notation easiness, it will be labeled as $y^2$;

- Real Gross Domestic Product (measured in billions of chained 2017 dollars), released by the U.S. Bureau of Economic Analysis. This series was used only for quarterly data estimations. Its notation will be $y^3$.

The sample period ranges between 1976/12 (1976Q4) and 2024/12 (2024Q4). As VAR models reflect short-run dynamics, I focus on the variables' cyclical components using the Hamilton filter (Hamilton, 2018). This trend-cycle decomposition technique only has not been used on $y$ and $y^2$, as these already express cyclical deviations from trend economic activity. Figure 1 displays monthly time-series plots of $y$, $e$, and $pr$. The latter two series are shown in natural logarithms and unfiltered.

[FIGURE 1 ABOUT HERE]

Finally, time-varying VAR models allow the researcher to compute, among other measures, impulse-response functions at any point in time within the sample period. I investigate periods of high and low growth using the National Bureau of Economic Research's (NBER) official business-cycle dating. Thus, the sample period covers six business-cycle peaks: 1980/01 (1980Q1), 1981/07 (1981Q3), 1990/07 (1990Q3), 2001/03 (2001Q1), 2007/12 (2007Q4), and 2020/02 (2019Q4); and six business-cycle troughs: 1980/07 (1980Q3), 1982/11 (1982Q4), 1991/03 (1991Q1), 2001/11 (2001Q4), 2009/06 (2009Q2), 2020/04 (2020Q2). The first dates represent high-, while the second denote low-growth periods.

Table 1 shows the mean and variance of $y$, $e$, and $pr$ for the five peak-to-peak business-cycle periods in the sample. No variable shows constant mean and variance: average energy use increased over time, diminishing only in the most recent period. A reverse pattern occurs with energy prices, with its second-highest mean value in the 2007/12–2020/02 interval. Their variances follow different patterns as well. While the former's variance decreased after the beginning of the 2000s, the latter's highest values were in the second and the last sub-periods. The economic activity index varies between positive and negative average values, with increasing variance over the cycles.

[TABLE 1 ABOUT HERE]

As a complement to the previous graph and table, Table 2 displays the average quarterly growth rates for energy use, real GDP, and real energy prices over the five business cycles analyzed in this paper.[6] The 1981Q3–1990Q3 and 1990Q3–2001Q1 business cycles show a combination of falling energy prices, increased energy use and, consequently, the highest average economic activity growth in the entire sample period. The most recent cycle, 2007Q4–2020Q2, while showing falling energy prices, also features decreasing energy use.

[TABLE 2 ABOUT HERE]

### 3.3 Identification

The identification of a reduced-form VAR model precedes any computational work. When the interest specifically lies in investigating impulse-response functions, a critical aspect involves the ordering of the endogenous variables vector. Variable ordering implies an *a priori* causal effect chain from one variable to the other(s) in the system. This ordering involves the use of orthogonal responses, adding the assumption that shocks do not occur in isolation but rather have contemporaneous effects on one, and lagged impact on other variable(s) in the system. At the same time, each shock is contemporaneously uncorrelated with other shocks.

The main empirical research question in the economic growth-energy use nexus literature lies in estimating its causal nature (Kraft and Kraft, 1978; Payne, 2009; Zhang and Cheng, 2009; Ajmi et al., 2015). However, no consensus exists, and it is likely that this paradigm will remain unchanged given the available data and standard econometric methods (Ozturk, 2010; Bruns et al., 2014). Although this study's interest drifts away from causality, properly identifying a VAR model implies establishing a recursive ordering of the endogenous variables. To that end, Bruns et al. (2014) show that models controlling for energy prices tend to offer a robust genuine causal relation from economic activity to energy use.

With this last point in mind, my approach is to estimate a parsimonious model where the rebound effect can be estimated without significant bias from omitted variables. At the same time, the goal is to introduce time-varying VAR models to this literature avoiding an overly-identified model. Thus, as energy demand is affected by the state of the economy as well as its prices, a three-variable model in energy use, a measure of economic activity, and a *proxy* for energy prices appears as an appropriate baseline selection.

Bruns et al. (2021), although not motivating their model using theoretical priors, utilize Independent Component Analysis (ICA) to find that the most stable recursive identification is $\mathbf{y}_t = (y, e, pr)'$. This implies that real economic activity contemporaneously affects (i.e., at time *t*) energy use, while the reverse is not true. Furthermore, energy use contemporaneously impacts real energy prices, but not vice-versa. In other words, energy prices lead energy use, and the latter leads real economic activity.[7]

---

[6]I omit growth rates for the other two economic activity variables on purpose. In terms of growth rates, it does not make intuitive sense to discuss growth rates of variables which are measured as standard deviations from its trend and ($y$) and as an index of percentage changes ($y^2$).

Given the lack of a consistent theory on the growth-energy interplay, I compute the cross-correlation coefficient function (CCF) among the three baseline variables to capture potential leading/lagging patterns in the utilized series.[8] Figure 2 plots the CCF between economic activity and energy use (top) and between energy use and its prices (bottom panel). The negative values on the horizontal axis represent lagging months, while positive numbers stand for leading periods. With the highest cross-correlation coefficients happening in lagging months, the panels reinforce the proposed recursive ordering: (*i*) energy use leads economic activity, and (*ii*) prices lead energy use.

[FIGURE 2 ABOUT HERE]

## 4 Results

This section presents and discusses the empirical estimates of the economy-wide rebound effect using the VAR methodology from Section 3. VAR models with time-varying parameters and stochastic volatility are efficiently estimated with Bayesian approaches. Denoting an unknown vector of time-varying parameters by $\mathbf{x}_t$ and recalling the vector of observed endogenous variables $\mathbf{y}_t = (y, e, pr)'$, the former can be estimated conditional on the latter through Bayes' theorem:

$$P(\mathbf{x}_t|\mathbf{y}_t) = \frac{f(\mathbf{y}_t|\mathbf{x}_t)P(\mathbf{x}_t)}{\int f(\mathbf{y}_t|\mathbf{x}_t)P(\mathbf{x}_t)d\mathbf{x}_t} \quad , \tag{10}$$

where $P(\mathbf{x}_t|\mathbf{y}_t)$ is the posterior density of $\mathbf{x}_t$; the right-hand side's numerator is the product of the likelihood function for $\mathbf{y}_t$ and any prior belief on the unknown vector $\mathbf{x}_t$; and the denominator is the normalizing constant. As both the likelihood function and the normalizing constant are usually intractable in Bayesian VAR settings, Markov Chain Monte Carlo (MCMC) algorithms recursively sample from the posterior distribution to provide an empirical estimate of the unknown parameter(s). Here, the unknown is the vector of responses of energy use to an energy efficiency shock at different points in time that make up the economy-wide rebound effect.

Recalling Equation (1)'s definition for the economy-wide rebound effect, it can be rewritten in the context of impulse-response analysis as follows:

---

[7]In this context, the term "leads" implies temporal precedence.
[8]See Shumway and Stoffer (2016) for a detailed overview of the cross-correlation coefficient function.

$$\text{Rebound effect}_i\ (\%) = \left[1 - \frac{x_i}{x_0}\right] \times 100, \tag{11}$$

where $x_0$ is the immediate response of energy use to an energy efficiency shock—thus denoting the expected percentage change in energy use—, and $x_i$ is the same response after $i$ periods, representing the actual change in energy demand.[9]

I utilize the same algorithm as in Del Negro and Primiceri (2015) to estimate all VAR models.[10] The MCMC procedure to compute the posterior distributions of all energy use responses was based on 55,000 draws, with the first 5,000 used as "burn-in" steps. Models with monthly and quarterly data were estimated with lag orders ($p$) 3 and 2, respectively, as suggested by the Akaike (AIC) and Schwarz (BIC) information criteria. Lastly, all models were computed with energy use and real energy prices in natural logarithms.

## 4.1 Baseline model

Following other VAR approaches in the literature, the energy efficiency innovation is represented by a negative, one-standard deviation shock to energy use. In order to compute the rebound effect, I focus on energy use's response to this latter shock over a 5-year horizon. As the MCMC procedure estimates an entire distribution (rather than single-point estimates) of the parameters of interest, my visualization approach proceeds in two steps.

First, I present energy use's posterior responses to an energy efficiency shock at 0 (immediate), 12 (4), 24 (8), 36 (12), 48 (16), and 60 (20) months (quarters) ahead for each business-cycle peak and trough period. In addition to the median posterior response (denoted by a solid black dot), I include the entire posterior response density (highlighting the underlying 66% density region), in order to incorporate the inherent uncertainty of Bayesian methods. Finally, I connect each horizon's median posterior response with a solid line to illustrate the responses' temporal evolution.

Second, I apply Equation (11) to the estimated posterior shock reactions, computing the percentage change between responses at time 0 and the subsequent years ahead for each business-cycle peak and trough period. These constitute the posterior, time-varying economy-wide rebound effects. This way, one can observe how their values change over time both in magnitude and volatility. The rebound effects are also shown in their posterior densities.

---

[9]See Bruns et al. (2021, p. 4) for a visual illustration of the economy-wide rebound effect using an impulse-response chart.

[10]See Nakajima (2011) and Krueger (2015) for detailed descriptions of the model's prior distributions and algorithmic specifications.

This subsection covers the baseline specification, with endogenous variables $y$, $e$, and $pr$ in monthly and quarterly frequency. The full impulse-response charts convey the expected signs and are shown in Online Appendix A, while here I focus solely on the graphs used for computing the economy-wide rebound effect.[11] Figure 3 displays the baseline model's impulse-response functions in monthly frequency for the six business-cycle peak periods in the sample. All reactions conform to the predicted sign, as energy use is expected to decrease after an energy-saving innovation. After the third year, however, posterior responses in all sub-periods start to approach zero. The strongest immediate responses to the shock occur in the first two sub-periods, 1980/01 and 1981/07, while the weakest occur in the two middle peaks, 1990/07 and 2001/03.

[FIGURE 3 ABOUT HERE]

Figure 4 presents energy use's posterior responses at quarterly frequency. Overall, posterior median response values and variances are smaller relative to those estimated in monthly frequency. While monthly posterior responses approach zero by the third year, quarterly median estimates stay below zero in most years. This pattern changes slightly in the last peak, 2019Q4, when median responses after 2 and 3 years are close to zero, decreasing over the next two years.

[FIGURE 4 ABOUT HERE]

Similar posterior values and patterns are observed in periods of low growth. Figure 5 presents the posterior responses of energy use to an energy efficiency shock across the six business-cycle trough months in the sample. Relative to peak periods, the immediate responses of energy use are stronger in magnitude in the first two sub-periods, 1980/07 and 1982/11. Quarterly troughs, displayed in Figure 6, closely resemble peak quarters, with exceptions at the second and fifth troughs, 1982Q4 and 2009Q2, with larger posterior median responses relative to the corresponding peaks. One important detail for quarterly estimations covering trough periods is that the MCMC algorithm was not able to estimate impulse-response functions for the last quarter, 2020Q2. Therefore, charts and tables for quarterly troughs will have one less panel/column.

[FIGURES 5 AND 6 ABOUT HERE]

After presenting the baseline models' impulse-response charts in monthly and quarterly frequency, I proceed to estimating the time-varying economy-wide rebound effect for each business-cycle peak and trough. Figure 7 starts off with the baseline model's posterior economy-wide rebound effects in a five-year interval for all business-cycle peak months. These are also shown in their entire posterior densities, with a dashed vertical line drawn at the 100% rebound mark.

---

[11]Economic activity decreases after negative, one-standard deviation exogenous shocks to energy use and real energy prices; energy use adjusts downward to a negative shock in economic activity after two years and positively to a fall in energy prices; and energy prices increase with negative disturbances both in economic activity and energy demand.

[FIGURE 7 ABOUT HERE]

To assist with the visual representation, Table 3 reports posterior median rebound effects at each year following the efficiency innovation. The only periods with median rebound values greater than 100%—implying a net increase (backfire) in energy use—are 3, 4, and 5 years after an energy-saving improvement in 1990/07. All other median values in the different business-cycle peaks lie below but close to the 100% threshold. The last two sub-periods, 2007/12 and 2020/02, show the largest variances among all peaks, denoted by their wider densities.

[TABLE 3 ABOUT HERE]

Rebound charts for peak quarters are displayed in Figure 8 and summarized in Table 4. Except for a median rebound of 100.3% two years after an energy efficiency shock in 2019Q4, all estimated posterior median rebounds do not indicate backfire, although lying close to the 100% mark. As observed in Bruns et al. (2021), models with quarterly data should estimate relatively smaller rebound effects for values lower than 100%, and this is what happens here as well.

[FIGURE 8 ABOUT HERE]

[TABLE 4 ABOUT HERE]

For the sake of compactness, all rebound charts for the baseline model's trough periods will be left to Online Appendix B, since these visually resemble peak months and quarters. Tables 5 and 6 show similar results to peak months and quarters, respectively.

[TABLES 5 AND 6 ABOUT HERE]

Results from baseline estimations suggest that almost the entirety of energy savings are put back into the system via increased energy use after approximately three years. In comparison, Bruns et al. (2021) estimate an overall rebound effect approaching 100% after four years in both monthly and quarterly frequencies, while Berner et al. (2022) find evidence of backfire after two years. The present time-varying analysis shows that backfire happens at specific time periods in the sample rather than being a permanent feature. More specifically, monthly estimations for business-cycle peaks and troughs show that the 1990s were the period of largest economy-wide rebound effects, with net increases in energy use reached after 3 years, slightly decreasing to 100% in the fifth year. Quarterly estimations, on the other hand, convey relatively smaller rebound effects, as previously suggested in the literature. In general, models using data in lower frequency estimate economy-wide rebound effects reaching their largest posterior values in the third year, decreasing thereafter.

It is not surprising to see the 1990s as the period with the largest rebound effect. As already shown in Tables 1 and 2, this cycle continued to experience falling energy prices and increased energy use from the previous decade, a combination that fueled the highest economic growth rates of the entire sample period. The 2000s, on the other hand, show a stagnation in the use of energy along with lower economic growth rates resulting in lower rebound effects. The baseline estimations show that economic growth utilizes energy efficiency gains, where while less energy is used to produce one more additional unit of output, more energy is demanded overall.

## 4.2 Alternative specifications

This subsection brings results for additional monthly and quarterly estimations. The baseline model was modified by using the alternative economic activity measures presented in Section 3, $y^2$ and $y^3$. While $y^2$ was used in both monthly and quarterly models, $y^3$ was only used in quarterly estimations. In order to present and discuss these results in a more compact way, all visualizations are available in the Online Appendix (Sections C and D). Here, I concentrate on the summary tables for peak months and quarters, as there is no considerable change to the results in low-growth periods.

Table 7 shows the economy-wide rebound effect posterior estimates using $y^2$ in monthly frequency. Results are similar to the baseline model's, with backfire rebound levels observed in the fourth and fifth years following an energy saving innovation in 1990/07. Tables 8 and 9 show similar posterior values to the baseline quarterly model, differing in that rebound values reach their peak in the fifth and not the third year after the energy efficiency shock.

[TABLES 7, 8, AND 9 ABOUT HERE]

From all estimations, the time-varying methodology shows that the 1990s were the period of largest economy-wide rebound effects in the U.S. economy. In models with monthly data, there is evidence of backfire, with a net increase in energy use after an energy-saving innovation. For estimations in quarterly frequency, rebound median values stay below the 100% mark. Over time, the economy-wide rebound effect has stayed close to 100%, implying that energy efficiency gains, while still saving a small amount of energy, are an insufficiently reliable instrument for the reduction of greenhouse gas emissions.

## 5 Conclusion

This work proposed a novel way of evaluating the macroeconomic dimensions of the energy rebound effect. The aim was to empirically estimate the endogenous time variations in energy use responding to energy efficiency innovations across periods of high and low growth. Vector Autoregressive (VAR) models have recently been implemented to this end in the empirical energy economics literature as an alternative to Computable General Equilibrium (CGE) models, as they avoid several parameterization issues that may influence final results and policy recommendations. To the best of my knowledge, this was the first application of time-varying VAR models to estimate the size of the economy-wide rebound effect.

Works such as Bruns et al. (2021) and Berner et al. (2022) innovate by computing the economy-wide rebound effect using impulse-response analysis. One limitation, however, of standard VAR methods is that these deliver single-point estimates which are supposed to describe dynamic interactions between two or more time series over an entire sample period. The present work introduced the time-varying VAR methodology with stochastic volatility from Primiceri (2005) and Del Negro and Primiceri (2015) to estimate the macroeconomic rebound effect across six business-cycle peaks: 1980/01 (1980Q1), 1981/07 (1981Q3), 1990/07 (1990Q3), 2001/03 (2001Q1), 2007/12 (2007Q4), and 2020/02 (2019Q4); and six business-cycle troughs: 1980/07 (1980Q3), 1982/11 (1982Q4), 1991/03 (1991Q1), 2001/11 (2001Q4), 2009/06 (2009Q2), 2020/04 (2020Q2). The first dates represent high-, while the second denote low-growth periods.

Time-varying VAR models with economic activity, energy use, and real energy prices as endogenous variables estimated an economy-wide rebound effect below but close to 100% in most business-cycle peak and trough periods. Monthly estimations for business-cycle peaks and troughs show that the 1990s were the period of largest economy-wide rebound effects, with backfire median values reached 3 years after an energy efficiency shock. Models in quarterly frequency, on the other hand, estimate relatively smaller rebound effects, analogous to other VAR applications to this phenomenon. The 1990s continued to experience falling energy prices and increased energy use, a combination that, on one hand, sustained high rates of economic activity that, on the other, neutralized the energy efficiency gains of the first studied cycles.

Overall, reductions in energy use following energy efficiency improvements last no longer than 3 years, implying that solely relying on energy-saving technological change has limited, short-lasting effects for reducing energy demand. Consequently, an effective departure from historical rebound and emission trends requires a diverse set of measures controlling for economic, institutional, and behavioral changes in energy use.

# 6  Bibliography

Adetutu, M. O., A. J. Glass, and T. G. Weyman-Jones (2016). Economy-wide estimates of rebound effects: Evidence from panel data. *The Energy Journal 37*(3), 251–270.

Ajmi, A. N., S. Hammoudeh, D. K. Nguyen, and J. R. Sato (2015). On the relationships between $CO_2$ emissions, energy consumption and income: The importance of time variation. *Energy Economics 49*, 629–638.

Ali, S., M. S. Ijaz, and I. Yousaf (2023). Dynamic spillovers and portfolio risk management between defi and metals: Empirical evidence from the Covid-19. *Resources Policy 83*, 103672.

Allan, G., N. Hanley, P. McGregor, K. Swales, and K. Turner (2007). The impact of increased efficiency in the industrial use of energy: A computable general equilibrium analysis for the United Kingdom. *Energy Economics 29*(4), 779–798.

Anson, S. and K. Turner (2009). Rebound and disinvestment effects in refined oil consumption and supply resulting from an increase in energy efficiency in the Scottish commercial transport sector. *Energy Policy 37*(9), 3608–3620.

Azevedo, I. M. (2014). Consumer end-use energy efficiency and rebound effects. *Annual Review of Environment and Resources 39*(1), 393–418.

Barker, T., A. Dagoumas, and J. Rubin (2009). The macroeconomic rebound effect and the world economy. *Energy Efficiency 2*, 411–427.

Berner, A., S. Bruns, A. Moneta, and D. I. Stern (2022). Do energy efficiency improvements reduce energy use? Empirical evidence on the economy-wide rebound effect in Europe and the United States. *Energy Economics 110*, 105939.

Brockway, P. E., H. Saunders, M. K. Heun, T. J. Foxon, J. K. Steinberger, J. R. Barrett, and S. Sorrell (2017). Energy rebound as a potential threat to a low-carbon future: Findings from a new exergy-based national-level rebound approach. *Energies 10*(1), 51.

Brockway, P. E., S. Sorrell, G. Semieniuk, M. K. Heun, and V. Court (2021). Energy efficiency and economy-wide rebound effects: A review of the evidence and its implications. *Renewable and Sustainable Energy Reviews 141*, 110781.

Brookes, L. (1990). The greenhouse effect: The fallacies in the energy efficiency solution. *Energy Policy 18*(2), 199–201.

Bruns, S. B., C. Gross, and D. I. Stern (2014). Is there really Granger causality between energy use and output? *The Energy Journal 35*(4), 101–134.

Bruns, S. B., A. Moneta, and D. I. Stern (2021). Estimating the economy-wide rebound effect using empirically identified structural vector autoregressions. *Energy Economics 97*, 105158.

Carriero, A., T. E. Clark, and M. Marcellino (2018). Measuring uncertainty and its impact on the economy. *Review of Economics and Statistics 100*(5), 799–815.

Clark, T. E. and F. Ravazzolo (2015). Macroeconomic forecasting performance under alternative specifications of time-varying volatility. *Journal of Applied Econometrics 30*(4), 551–575.

Cogley, T. and T. J. Sargent (2001). Evolving post-World War II US inflation dynamics. *NBER Macroeconomics Annual 16*, 331–373.

Del Negro, M. and G. E. Primiceri (2015). Time varying structural vector autoregressions and monetary policy: A corrigendum. *The Review of Economic Studies 82*(4), 1342–1345.

Duarte, R., J. Sánchez-Chóliz, and C. Sarasa (2018). Consumer-side actions in a low-carbon economy: A dynamic CGE analysis for Spain. *Energy Policy 118*, 199–210.

Galí, J. and L. Gambetti (2015). The effects of monetary policy on stock market bubbles: Some evidence. *American Economic Journal: Macroeconomics 7*(1), 233–257.

Gillingham, K., D. Rapson, and G. Wagner (2016). The rebound effect and energy efficiency policy. *Review of Environmental Economics and Policy*.

Greening, L. A., D. L. Greene, and C. Difiglio (2000). Energy efficiency and consumption—the rebound effect—a survey. *Energy Policy 28*(6-7), 389–401.

Grubb, M. (1990). Communication energy efficiency and economic fallacies. *Energy Policy 18*(8), 783–785.

Guerra, A.-I. and F. Sancho (2010). Rethinking economy-wide rebound measures: An unbiased proposal. *Energy Policy 38*(11), 6684–6694.

Hamilton, J. D. (2018). Why you should never use the Hodrick–Prescott filter. *Review of Economics and Statistics 100*(5), 831–843.

Khazzoom, J. D. (1980). Economic implications of mandated efficiency in standards for household appliances. *The Energy Journal 1*(4), 21–40.

Kilian, L. (2009). Not all oil price shocks are alike: Disentangling demand and supply shocks in the crude oil market. *American Economic Review 99*(3), 1053–1069.

Kilian, L. (2019). Measuring global real economic activity: Do recent critiques hold up to scrutiny? *Economics Letters 178*, 106–110.

Kilian, L. and X. Zhou (2018). Modeling fluctuations in the global demand for commodities. *Journal of International Money and Finance 88*, 54–78.

Koop, G., D. Korobilis, et al. (2010). Bayesian multivariate time series methods for empirical macroeconomics. *Foundations and Trends in Econometrics 3*(4), 267–358.

Kraft, J. and A. Kraft (1978). On the relationship between energy and GNP. *The Journal of Energy and Development*, 401–403.

Krueger, F. (2015). bvarsv: Bayesian analysis of a vector autoregressive model with stochastic volatility and time-varying parameters. *R package version 1*.

Kulmer, V. and S. Seebauer (2019). How robust are estimates of the rebound effect of energy efficiency improvements? A sensitivity analysis of consumer heterogeneity and elasticities. *Energy Policy 132*, 1–14.

Lütkepohl, H. (2006). *New introduction to multiple time series analysis*. Springer.

Nakajima, J. (2011). Time-varying parameter VAR model with stochastic volatility: An overview of methodology and empirical applications. *Institute for Monetary and Economic Studies, Bank of Japan Tokyo, Japan*.

Ozturk, I. (2010). A literature survey on energy–growth nexus. *Energy Policy 38*(1), 340–349.

Payne, J. E. (2009). On the dynamics of energy consumption and output in the US. *Applied Energy 86*(4), 575–577.

Pfaff, B. (2008). *Analysis of integrated and cointegrated time series with R*. Springer Science & Business Media.

Primiceri, G. E. (2005). Time varying structural vector autoregressions and monetary policy. *The Review of Economic Studies 72*(3), 821–852.

Shumway, R. H. and D. S. Stoffer (2016). *Time series analysis and its applications with R examples* (Fourth ed.). Oxford.

Sims, C. A. (2001). Comment on Sargent and Cogley's 'Evolving US Postwar Inflation Dynamics'. *NBER Macroeconomics Annual 16*, 373–379.

Sorrell, S. and J. Dimitropoulos (2008). The rebound effect: Microeconomic definitions, limitations and extensions. *Ecological Economics 65*(3), 636–649.

Stock, J. H. and M. W. Watson (2007). Why has US inflation become harder to forecast? *Journal of Money, Credit and banking 39*, 3–33.

Turner, K. and N. Hanley (2011). Energy efficiency, rebound effects and the environmental Kuznets Curve. *Energy Economics 33*(5), 709–720.

U.S. Census Bureau (2023). X-13ARIMA-SEATS Reference Manual. https://www2.census.gov/software/x-13arima-seats/x13as/windows/documentation/docx13ashtml.pdf. Accessed on May 2nd, 2025.

U.S. International Energy Agency (2025). Energy Technology Perspectives 2024. https://www.iea.org/reports/energy-technology-perspectives-2024. Accessed on April 30th, 2025.

Wei, T. and Y. Liu (2017). Estimation of global rebound effect caused by energy efficiency improvement. *Energy Economics 66*, 27–34.

Wei, T., J. Zhou, and H. Zhang (2019). Rebound effect of energy intensity reduction on energy consumption. *Resources, Conservation and Recycling 144*, 233–239.

Zhang, X.-P. and X.-M. Cheng (2009). Energy consumption, carbon emissions, and economic growth in China. *Ecological Economics 68*(10), 2706–2712.

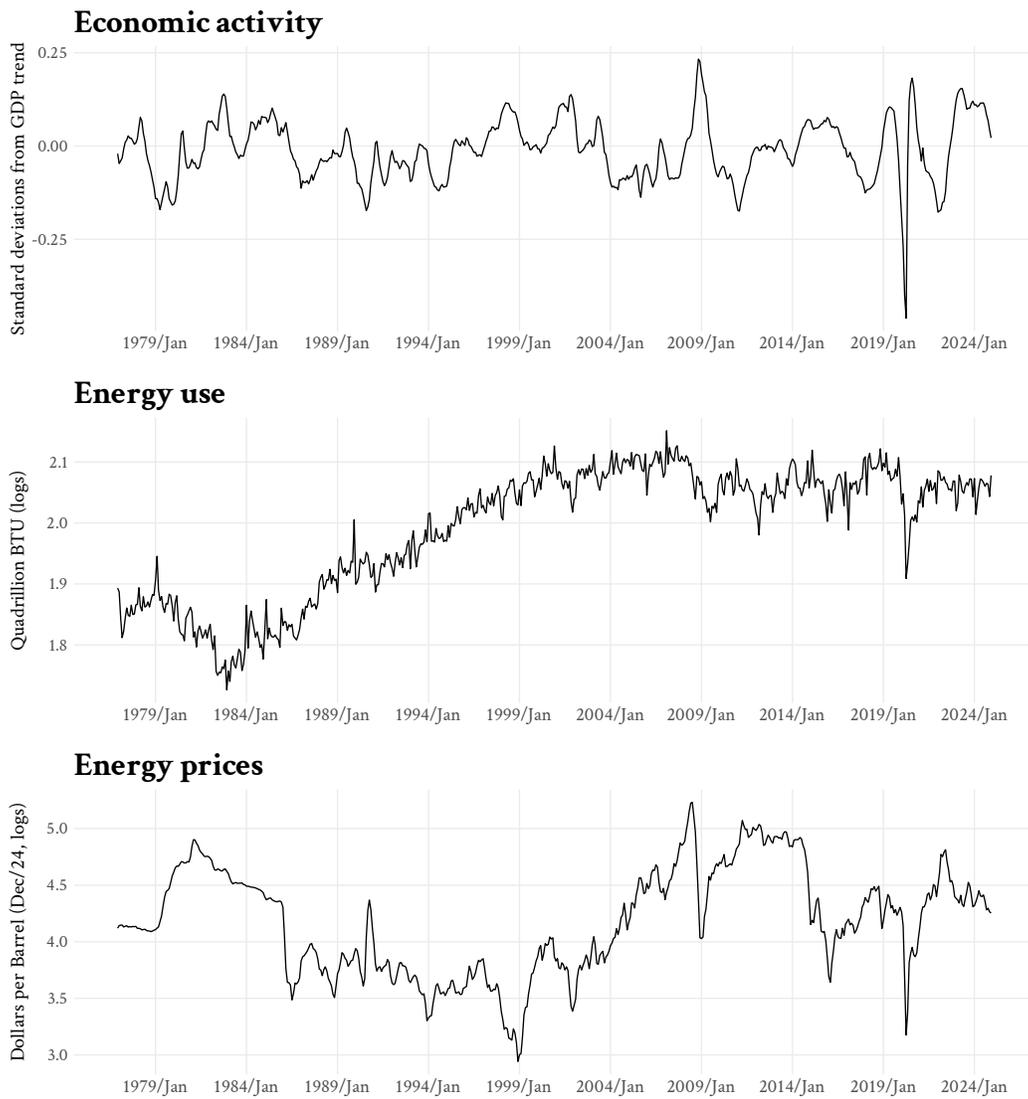

Figure 1: Time plots of U.S. economic activity (*y*, top), energy use (*e*, middle), and energy prices (*pr*, bottom), December/1976–December/2024. Note: Energy use and real prices are log-transformed.

Table 1: Mean and variance, baseline variables, peak-to-peak U.S. business cycles

|                   | Energy use |          | Economic activity |          | Energy prices |          |
|-------------------|------------|----------|-------------------|----------|---------------|----------|
| **Period**        | Mean       | Variance | Mean              | Variance | Mean          | Variance |
| 1980/01–1981/07   | 6.26       | 0.02     | -0.0530           | 0.0025   | 115.89        | 111.11   |
| 1981/07–1990/07   | 6.32       | 0.13     | 0.0020            | 0.0043   | 68.43         | 702.25   |
| 1990/07–2001/03   | 7.44       | 0.16     | -0.0128           | 0.0039   | 39.42         | 101.55   |
| 2001/03–2007/12   | 8.08       | 0.03     | -0.0289           | 0.0054   | 68.73         | 654.61   |
| 2007/12–2020/02   | 7.87       | 0.04     | -0.0069           | 0.0068   | 101.32        | 1287.53  |

Table 2: Average quarterly growth rates (%), peak-to-peak U.S. business cycles

| **Period**        | Energy use | Real GDP | Energy prices |
|-------------------|------------|----------|---------------|
| 1980/01–1981/07   | -0.705     | 0.328    | 2.15          |
| 1981/07–1990/07   | 0.302      | 0.820    | -3.85         |
| 1990/07–2001/03   | 0.311      | 0.806    | -1.44         |
| 2001/03–2007/12   | 0.064      | 0.649    | 3.23          |
| 2007/12–2020/02   | -0.165     | 0.408    | -3.00         |

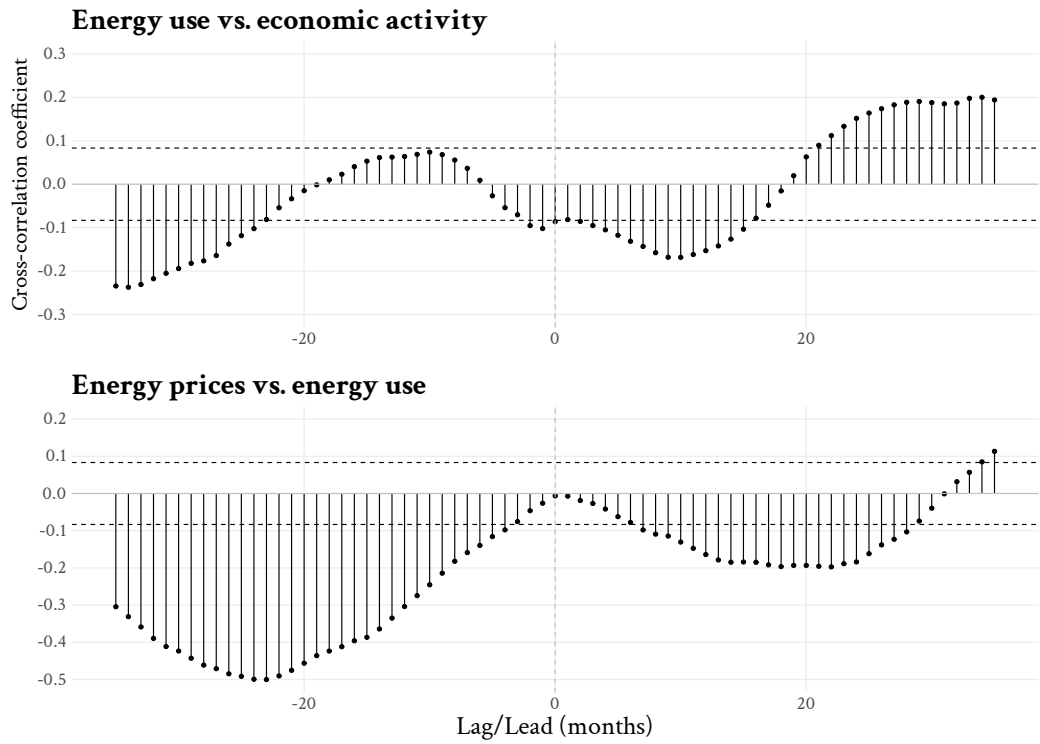

Figure 2: Monthly cross-correlation coefficient plots. Top: Energy use *vs.* economic activity; bottom: Energy prices *vs.* energy use. Dashed horizontal lines indicate $\pm 2 \times 1/\sqrt{n}$, with $n = 577$ observations.

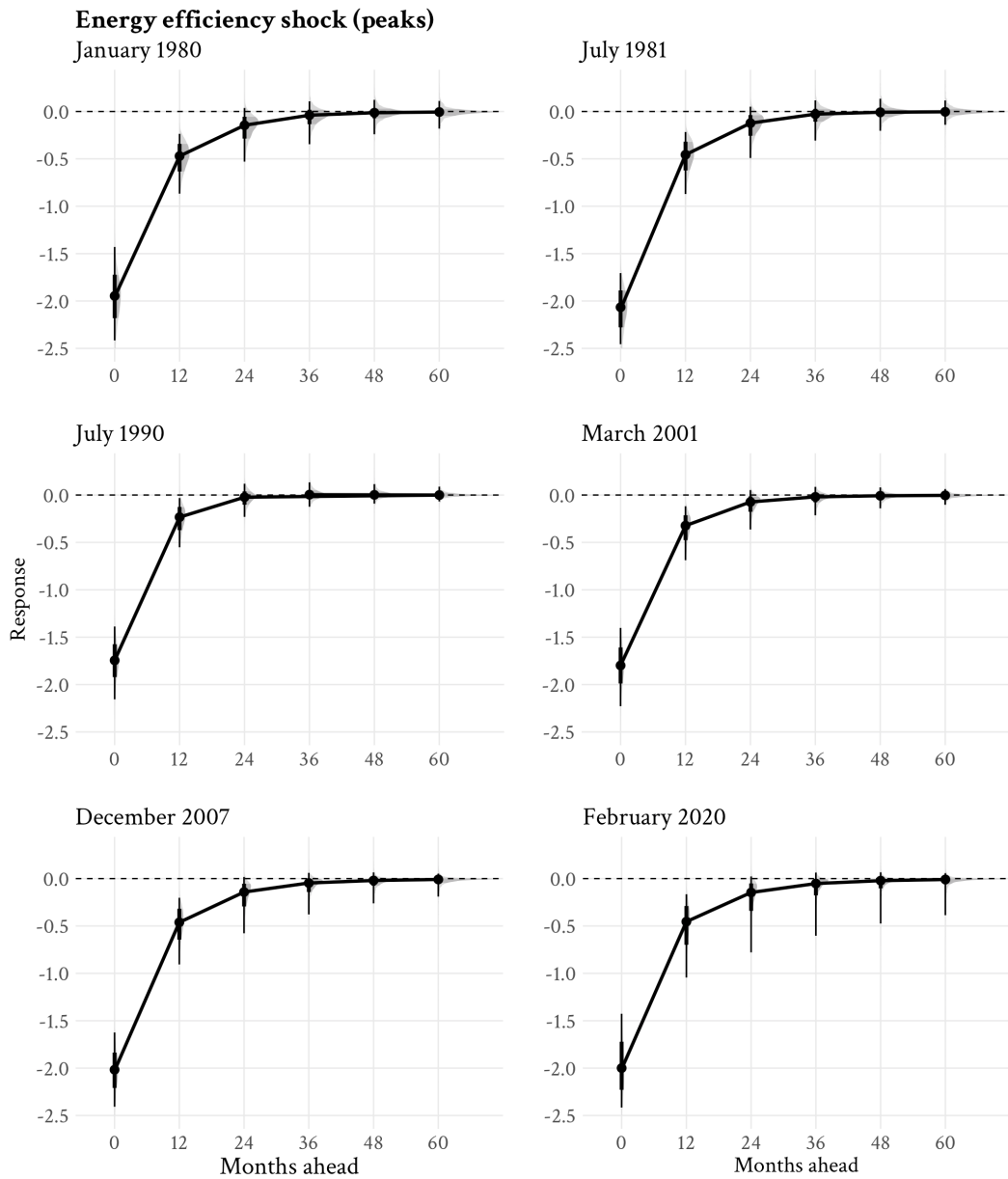

Figure 3: Posterior responses of energy use to a one-standard deviation shock in energy efficiency at different NBER-dated business-cycle peak months. Note: Black dots denote posterior medians; darker shaded areas mark the 66% posterior density region; and the solid line connects posterior medians.

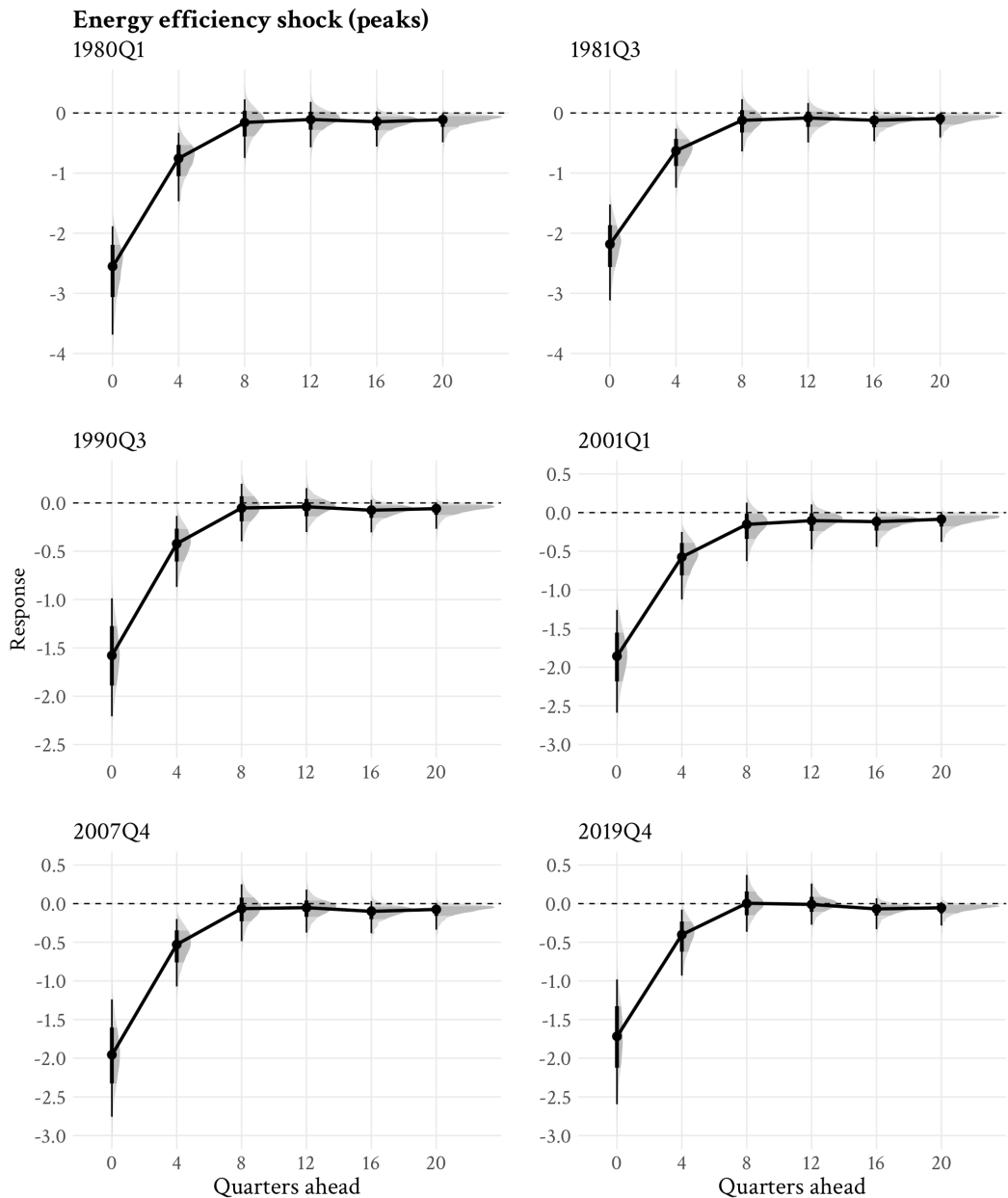

Figure 4: Posterior responses of energy use to a one-standard deviation shock in energy efficiency at different NBER-dated business-cycle peak quarters. Note: Black dots denote posterior medians; darker shaded areas mark the 66% posterior density region; and the solid line connects posterior medians.

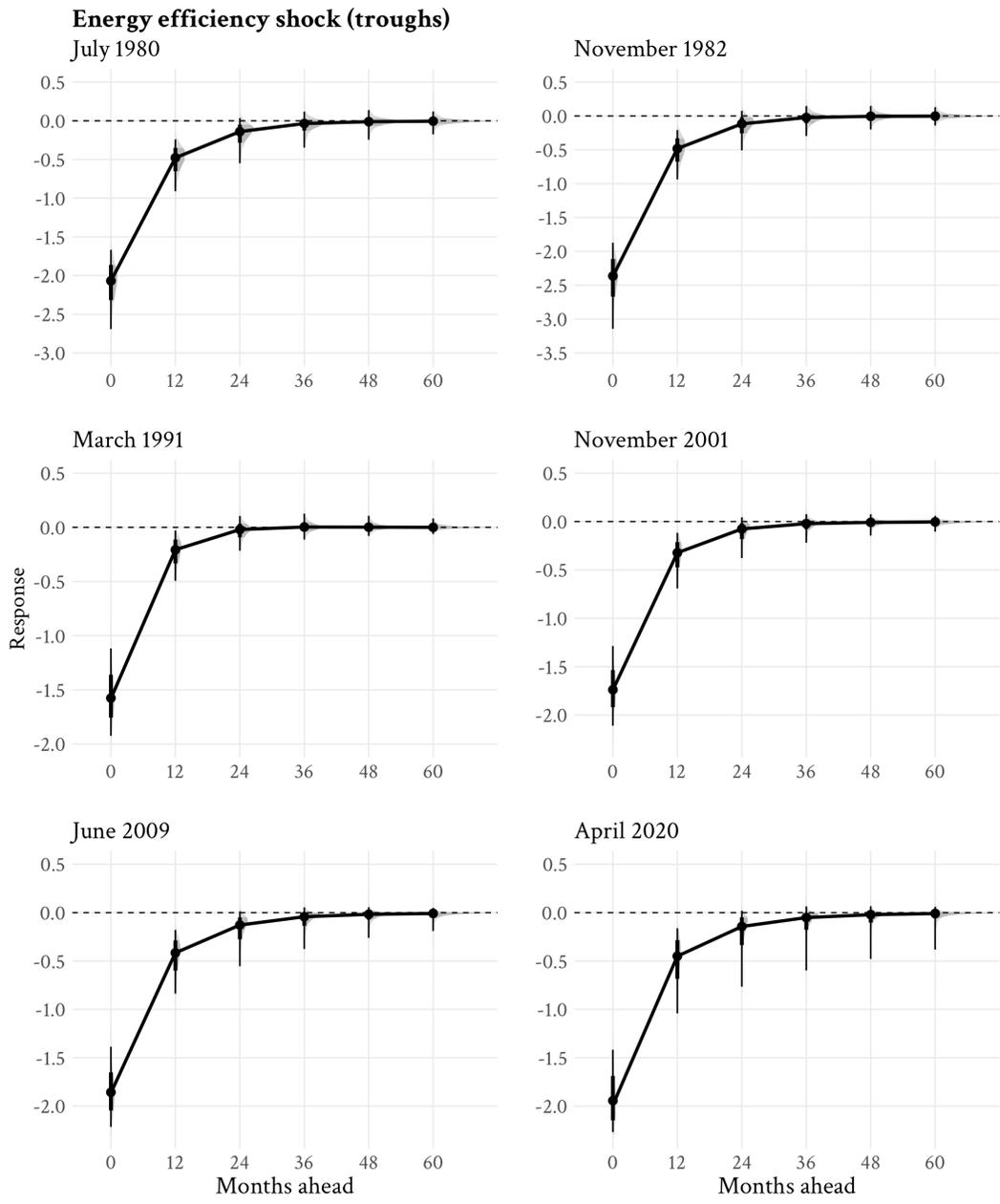

Figure 5: Posterior responses of energy use to a one-standard deviation shock in energy efficiency at different NBER-dated business-cycle trough months. Note: Black dots denote posterior medians; darker shaded areas mark the 66% posterior density region; and the solid line connects posterior medians.

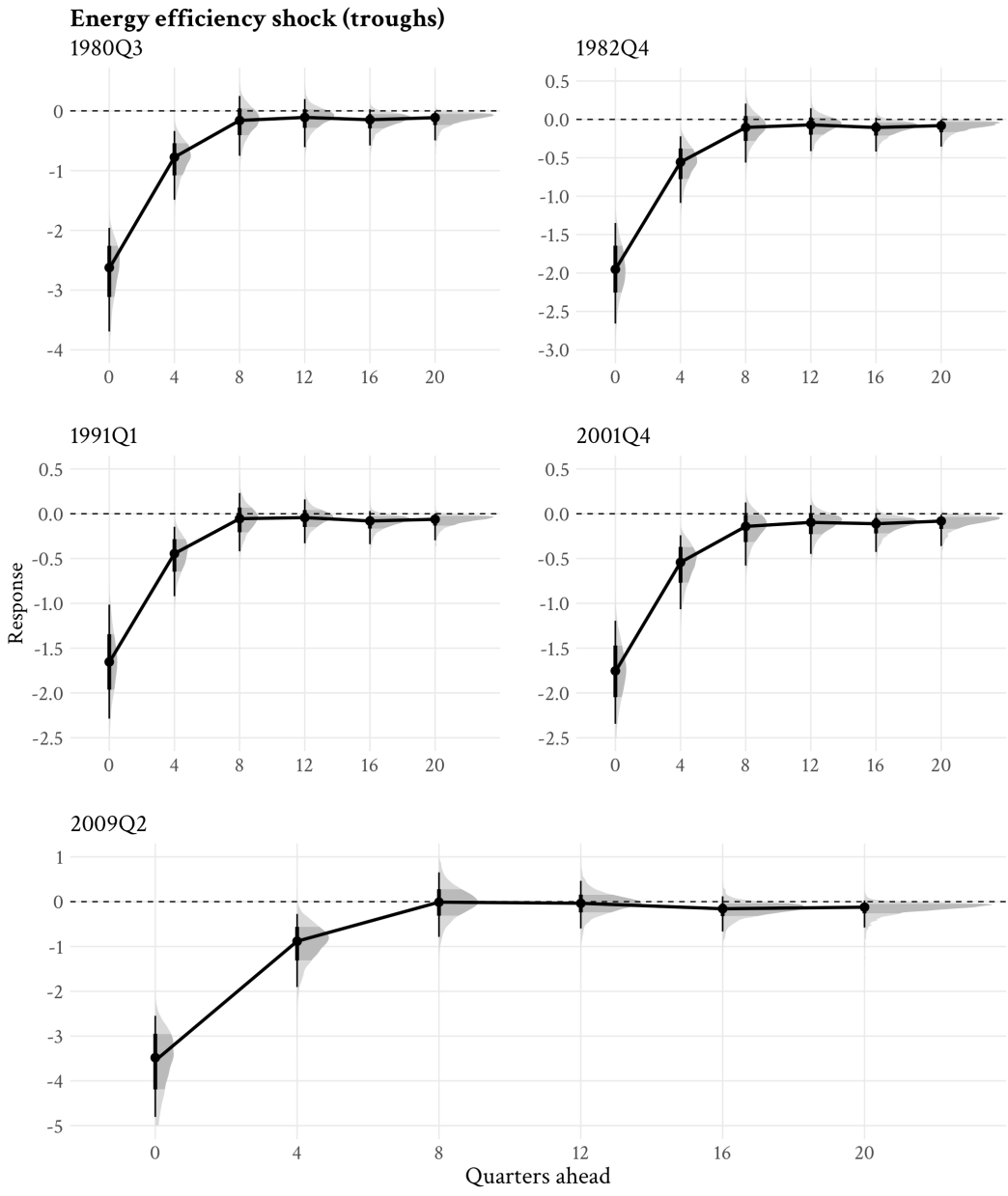

Figure 6: Posterior responses of energy use to a one-standard deviation shock in energy efficiency at different NBER-dated business-cycle trough quarters. Note: Black dots denote posterior medians; darker shaded areas mark the 66% posterior density region; and the solid line connects posterior medians.

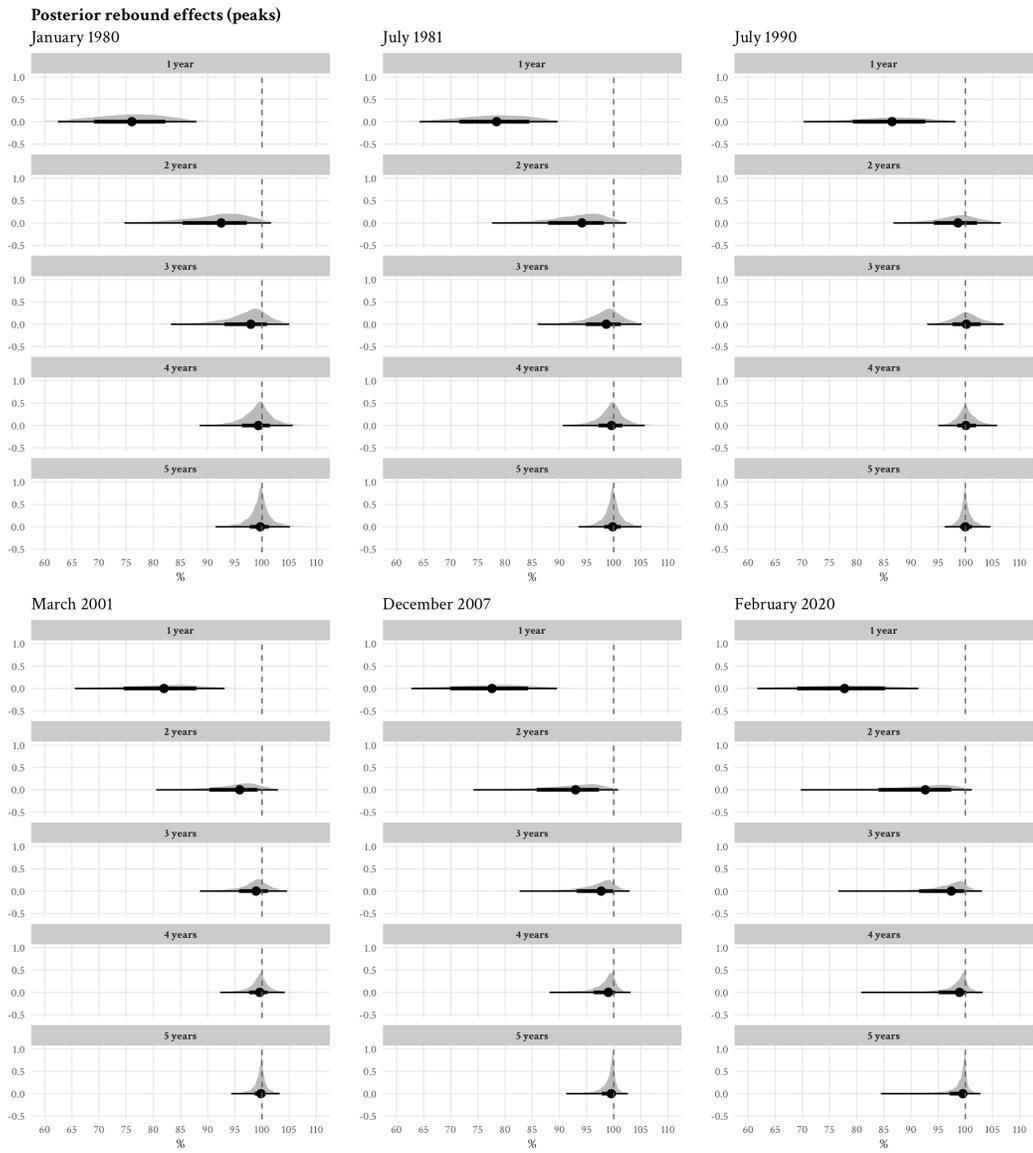

Figure 7: Posterior rebound effects at 1-, 2-, 3-, 4-, and 5-year intervals (business-cycle peak months). Note: Darker shaded areas denote the 90% density region.

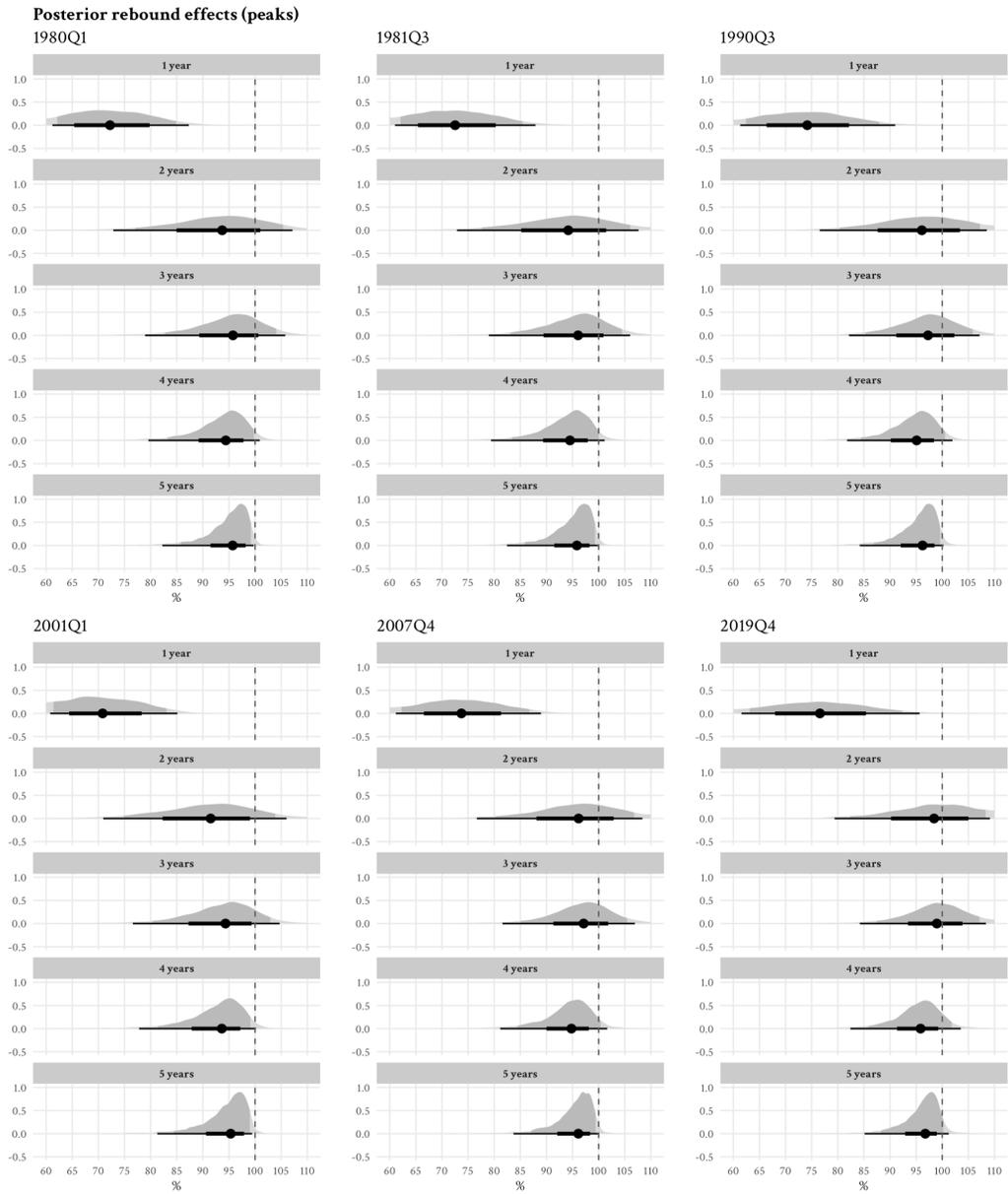

Figure 8: Posterior rebound effects at 1-, 2-, 3-, 4-, and 5-year intervals (business-cycle peak quarters). Note: Darker shaded areas denote the 90% density region.

Table 3: Posterior median rebound effect, U.S. business-cycle peak months, baseline model

| Horizon | 1980/01 | 1981/07 | 1990/07 | 2001/03 | 2007/12 | 2020/02 |
|---|---|---|---|---|---|---|
| 1 year  | 75.7 [65.4, 84.0]  | 78.2 [68.2, 86.2]  | 86.5 [76.1, 94.6]  | 81.9 [70.7, 89.8]  | 77.2 [65.5, 86.1]  | 76.9 [61.6, 87.2] |
| 2 years | 92.5 [82.3, 98.6]  | 94.1 [85.3, 99.5]  | 98.7 [92.1, 103.6] | 95.9 [87.6, 100.3] | 92.9 [82.2, 98.4]  | 92.5 [78.1, 98.6] |
| 3 years | 98.0 [90.7, 102.9] | 98.7 [92.7, 102.5] | 100.2 [96.5, 104.2]| 98.9 [94.0, 102.0] | 97.7 [90.4, 100.7] | 97.4 [87.0, 100.5] |
| 4 years | 99.3 [94.7, 102.8] | 99.6 [95.8, 102.9] | 100.1 [97.7, 103.2]| 99.6 [96.4, 101.9] | 99.0 [94.4, 100.8] | 98.9 [91.4, 100.8] |
| 5 years | 99.7 [96.6, 102.4] | 99.8 [97.2, 102.4] | 100.0 [98.4, 102.1]| 99.8 [97.7, 101.3] | 99.5 [96.5, 100.6] | 99.5 [94.4, 100.6] |

*Note*: $10^{th}$ and $90^{th}$ percentile values in square brackets.

Table 4: Posterior median rebound effect, U.S. business-cycle peak quarters, baseline model

| Horizon | 1980Q1 | 1981Q3 | 1990Q3 | 2001Q1 | 2007Q4 | 2019Q4 |
|---|---|---|---|---|---|---|
| 1 year  | 70.5 [57.5, 81.5]  | 70.9 [57.6, 82.0]  | 72.9 [59.8, 84.3]  | 68.5 [55.4, 79.6]  | 72.6 [59.8, 83.6]  | 75.8 [61.8, 88.3] |
| 2 years | 93.9 [81.4, 103.9] | 94.4 [81.5, 104.6] | 96.6 [84.5, 107.6] | 91.5 [78.4, 101.8] | 96.6 [84.7, 106.6] | 100.3 [88.1, 113.6] |
| 3 years | 95.8 [86.4, 102.7] | 96.1 [86.5, 103.0] | 97.4 [88.8, 104.8] | 94.4 [84.3, 101.2] | 97.3 [88.8, 104.1] | 99.4 [91.1, 107.9] |
| 4 years | 94.4 [86.6, 98.7]  | 94.5 [86.7, 98.8]  | 95.1 [87.9, 99.4]  | 93.6 [85.0, 98.2]  | 94.8 [87.4, 99.2]  | 95.9 [89.0, 100.5] |
| 5 years | 95.7 [89.1, 98.7]  | 95.8 [89.1, 98.8]  | 96.2 [90.0, 99.1]  | 95.3 [88.0, 98.5]  | 96.1 [89.8, 98.9]  | 96.8 [90.8, 99.5] |

*Note*: $10^{th}$ and $90^{th}$ percentile values in square brackets.

Table 5: Posterior median rebound effect, U.S. business-cycle trough months, baseline model

| Horizon | 1980/07 | 1982/11 | 1991/03 | 2001/11 | 2009/06 | 2020/04 |
|---|---|---|---|---|---|---|
| 1 year  | 76.9 [66.6, 84.8]  | 79.7 [69.5, 87.8]  | 86.4 [76.1, 94.4]  | 81.2 [70.0, 89.1]  | 77.3 [65.1, 86.0]  | 77.0 [61.8, 87.3] |
| 2 years | 93.2 [83.5, 98.9]  | 95.1 [86.6, 100.3] | 98.7 [92.2, 103.5] | 95.5 [86.8, 100.0] | 92.9 [81.7, 98.3]  | 92.6 [78.3, 98.6] |
| 3 years | 98.3 [91.5, 102.3] | 99.0 [93.7, 102.9] | 100.2 [96.4, 104.2]| 98.8 [93.6, 101.7] | 97.7 [90.2, 100.6] | 97.4 [87.0, 100.6] |
| 4 years | 99.5 [95.1, 102.8] | 99.8 [96.3, 102.9] | 100.1 [97.7, 103.2]| 99.5 [96.3, 101.7] | 99.1 [94.1, 100.8] | 98.9 [91.7, 100.8] |
| 5 years | 99.8 [96.9, 102.5] | 99.9 [97.5, 102.3] | 100.0 [98.5, 102.1]| 99.8 [97.7, 101.2] | 99.6 [96.4, 100.6] | 99.5 [94.4, 100.6] |

*Note*: $10^{th}$ and $90^{th}$ percentile values in square brackets.

Table 6: Posterior median rebound effect, U.S. business-cycle trough quarters, baseline model

| Horizon | 1980Q3 | 1982Q4 | 1991Q1 | 2001Q4 | 2009Q2 |
|---|---|---|---|---|---|
| 1 year  | 70.7 [57.4, 81.7]  | 71.2 [58.0, 82.2]  | 72.7 [59.6, 84.0]  | 68.6 [55.6, 79.7]  | 75.2 [62.0, 86.6 ] |
| 2 years | 93.9 [81.5, 104.2] | 94.6 [82.2, 104.8] | 96.5 [84.2, 107.4] | 91.8 [78.7, 101.9] | 99.7 [88.2, 110.9] |
| 3 years | 95.9 [86.4, 102.8] | 96.3 [87.0, 103.1] | 97.3 [88.4, 104.5] | 94.4 [84.6, 101.3] | 99.0 [91.2, 106.7] |
| 4 years | 94.4 [86.7, 98.8]  | 94.6 [87.0, 98.9]  | 95.0 [87.7, 99.3]  | 93.6 [85.2, 98.2]  | 95.7 [89.0, 100.0] |
| 5 years | 95.8 [89.1, 98.7]  | 95.8 [89.4, 98.8]  | 96.2 [89.9, 99.0]  | 95.3 [88.1, 98.5]  | 96.6 [91.0, 99.3] |

*Note*: $10^{th}$ and $90^{th}$ percentile values in square brackets.

Table 7: Posterior median rebound effect, U.S. business-cycle peak months, model with $y^2$

| Horizon | 1980/01 | 1981/07 | 1990/07 | 2001/03 | 2007/12 | 2020/02 |
|---|---|---|---|---|---|---|
| 1 year  | 76.0 [65.4, 83.8]  | 77.3 [67.1, 85.0]  | 83.2 [72.9, 91.0]  | 80.4 [69.5, 88.5]  | 78.2 [66.5, 86.4]  | 73.9 [57.7, 85.1] |
| 2 years | 91.4 [80.8, 97.1]  | 92.4 [82.8, 97.7]  | 96.4 [88.9, 101.0] | 94.5 [85.9, 99.0]  | 93.1 [83.0, 98.0]  | 91.0 [75.1, 97.9] |
| 3 years | 97.2 [89.0, 100.7] | 97.7 [90.9, 101.0] | 99.5 [94.9, 102.7] | 98.4 [93.0, 101.0] | 97.7 [91.0, 100.2] | 96.8 [85.0, 100.5] |
| 4 years | 99.2 [93.6, 101.8] | 99.4 [94.9, 102.0] | 100.1 [97.3, 102.5]| 99.4 [96.1, 101.4] | 99.1 [94.9, 100.6] | 98.7 [90.1, 100.9] |
| 5 years | 99.8 [96.2, 102.0] | 99.9 [97.0, 102.1] | 100.1 [98.3, 102.1]| 99.8 [97.5, 101.2] | 99.6 [96.9, 100.5] | 99.4 [93.2, 100.7] |

*Note*: $10^{th}$ and $90^{th}$ percentile values in square brackets.

Table 8: Posterior median rebound effect, U.S. business-cycle peak quarters, model with $y^2$

| Horizon | 1980Q1 | 1981Q3 | 1990Q3 | 2001Q1 | 2007Q4 | 2019Q4 |
|---|---|---|---|---|---|---|
| 1 year | 66.5 [52.1, 77.5] | 66.8 [52.7, 77.7] | 68.5 [54.3, 78.9] | 67.3 [53.4, 78.0] | 67.9 [54.1, 78.6] | 69.0 [55.4, 80.0] |
| 2 years | 88.9 [77.0, 95.9] | 89.3 [77.5, 96.1] | 90.4 [79.2, 97.1] | 89.6 [77.8, 96.1] | 90.1 [78.8, 96.8] | 90.9 [79.8, 98.0] |
| 3 years | 96.3 [88.2, 100.2] | 96.4 [88.5, 100.2] | 97.1 [89.8, 101.0] | 96.6 [88.8, 100.2] | 97.0 [89.7, 100.5] | 97.5 [90.4, 101.4] |
| 4 years | 98.6 [93.4, 100.9] | 98.7 [93.6, 101.0] | 99.1 [94.3, 101.4] | 98.8 [93.8, 100.9] | 99.0 [94.3, 101.1] | 99.2 [94.8, 101.7] |
| 5 years | 99.4 [95.8, 100.8] | 99.5 [95.9, 100.8] | 99.6 [96.5, 101.1] | 99.5 [96.1, 100.8] | 99.6 [96.4, 100.9] | 99.7 [96.7, 101.3] |

*Note*: $10^{th}$ and $90^{th}$ percentile values in square brackets.

Table 9: Posterior median rebound effect, U.S. business-cycle peak quarters, model with $y^3$

| Horizon | 1980Q1 | 1981Q3 | 1990Q3 | 2001Q1 | 2007Q4 | 2019Q4 |
|---|---|---|---|---|---|---|
| 1 year | 79.6 [69.5, 87.8] | 79.7 [69.6, 87.9] | 80.3 [70.3, 88.5] | 79.4 [69.5, 87.5] | 79.5 [69.2, 87.4] | 79.8 [69.4, 88.0] |
| 2 years | 92.6 [85.1, 98.4] | 92.8 [85.2, 98.5] | 93.4 [85.9, 99.2] | 92.4 [84.8, 98.0] | 92.3 [84.7, 98.0] | 92.8 [85.0, 98.6] |
| 3 years | 96.2 [90.0, 100.8] | 96.4 [90.1, 101.0] | 96.9 [90.8, 101.8] | 96.0 [89.7, 100.4] | 96.0 [89.7, 100.4] | 96.4 [89.9, 101.0] |
| 4 years | 97.8 [92.5, 101.6] | 98.0 [92.6, 101.7] | 98.4 [93.3, 102.5] | 97.7 [92.3, 101.2] | 97.6 [92.2, 101.1] | 98.0 [92.6, 101.8] |
| 5 years | 98.8 [94.2, 101.7] | 98.8 [94.2, 101.8] | 99.2 [95.0, 102.5] | 98.6 [94.1, 101.4] | 98.6 [93.9, 101.3] | 98.8 [94.3, 101.8] |

*Note*: $10^{th}$ and $90^{th}$ percentile values in square brackets.

# The economy-wide rebound effect and U.S. business cycles: A time-varying exercise
# Online Appendix

Marcio Santetti[*]

This document accompanies Santetti, M. (2025). "The economy-wide rebound effect and U.S. business cycles: A time-varying exercise." It includes four sections. The first details the full impulse-response function charts for the baseline model with quarterly data. The second displays the baseline model's rebound charts for trough months and quarters. The third provides impulse-response and rebound charts for the alternative specifications using different economic activity proxy variables. Finally, the last section brings additional summary tables with the size of the economy-wide rebound effect for the alternative model specifications.

**Keywords**: Economy-wide rebound effect; Energy use; Time-varying VAR models; Bayesian inference.

**JEL Classification**: C32, Q43

---

[*]Emerson College. Email: marcio.santetti@emerson.edu.

# A Full impulse-response functions charts (quarterly data)

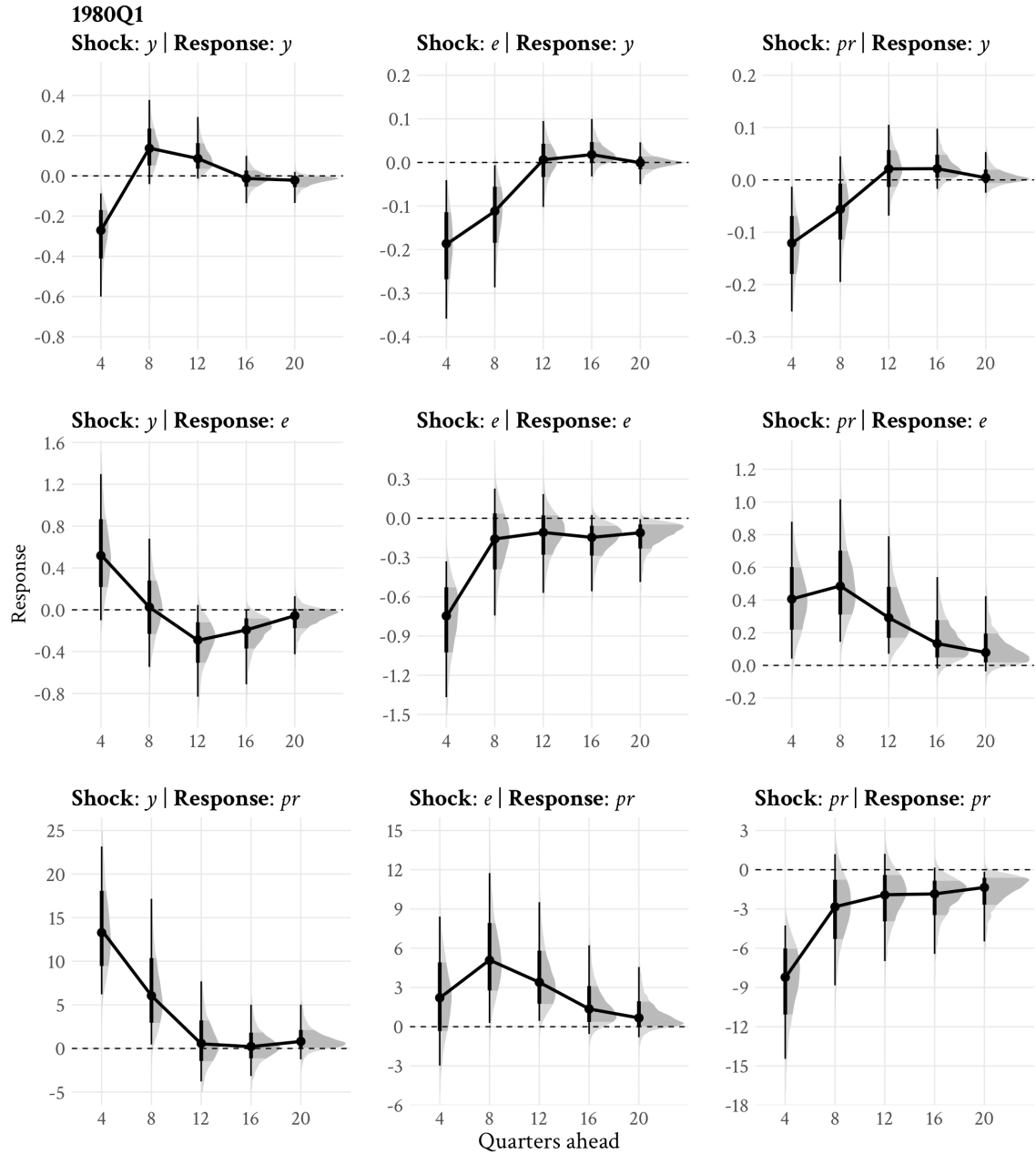

Figure 1: Posterior impulse-response functions in 1980Q1, baseline model. Note 1: All shocks denote a negative, one standard deviation exogenous innovation to the model variables. Note 2: Black dots denote posterior medians; darker shaded areas mark the 66% posterior density region; and the solid line connects posterior medians.

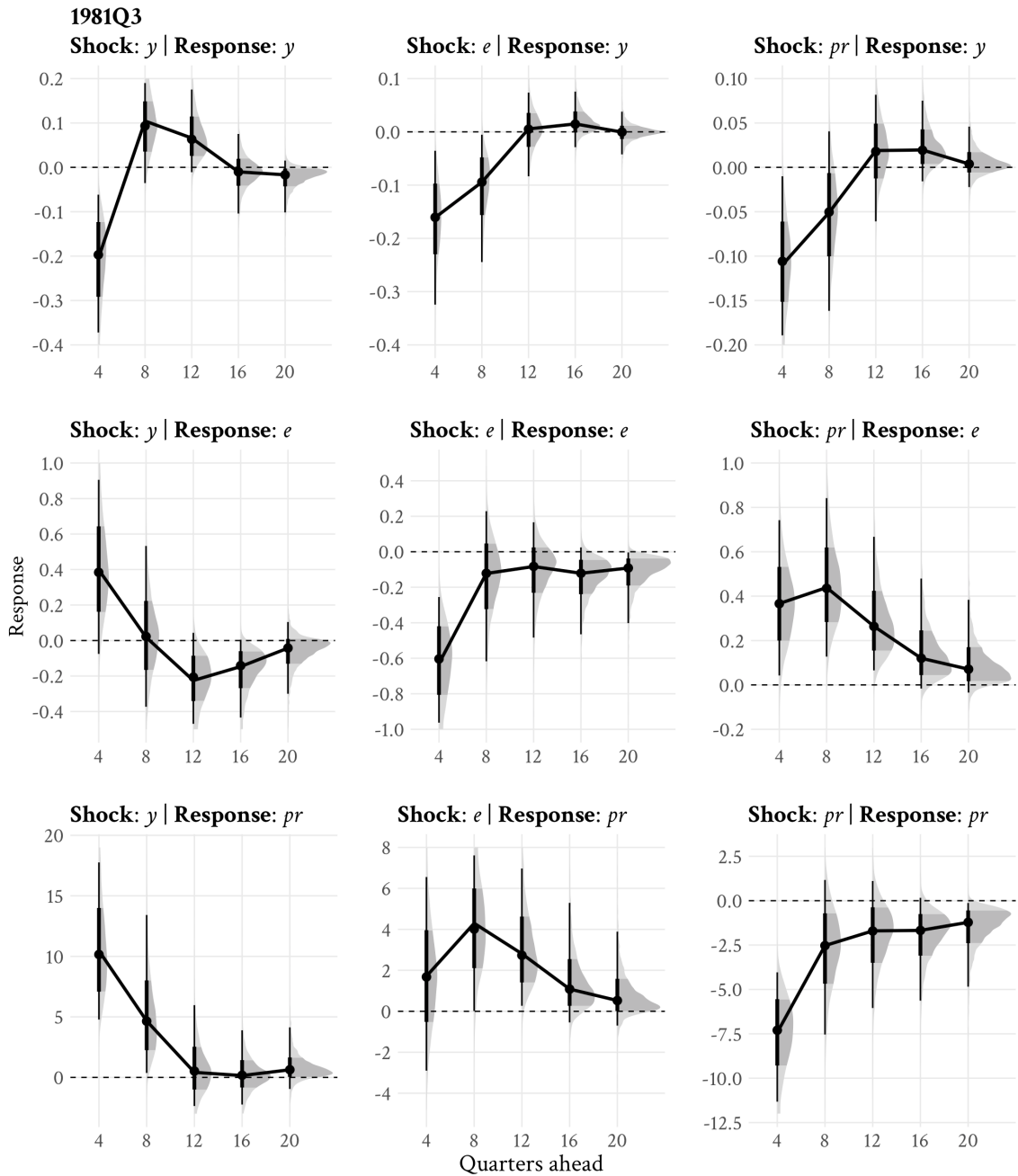

Figure 2: Posterior impulse-response functions in 1981Q3, baseline model. Note 1: All shocks denote a negative, one standard deviation exogenous innovation to the model variables. Note 2: Black dots denote posterior medians; darker shaded areas mark the 66% posterior density region; and the solid line connects posterior medians.

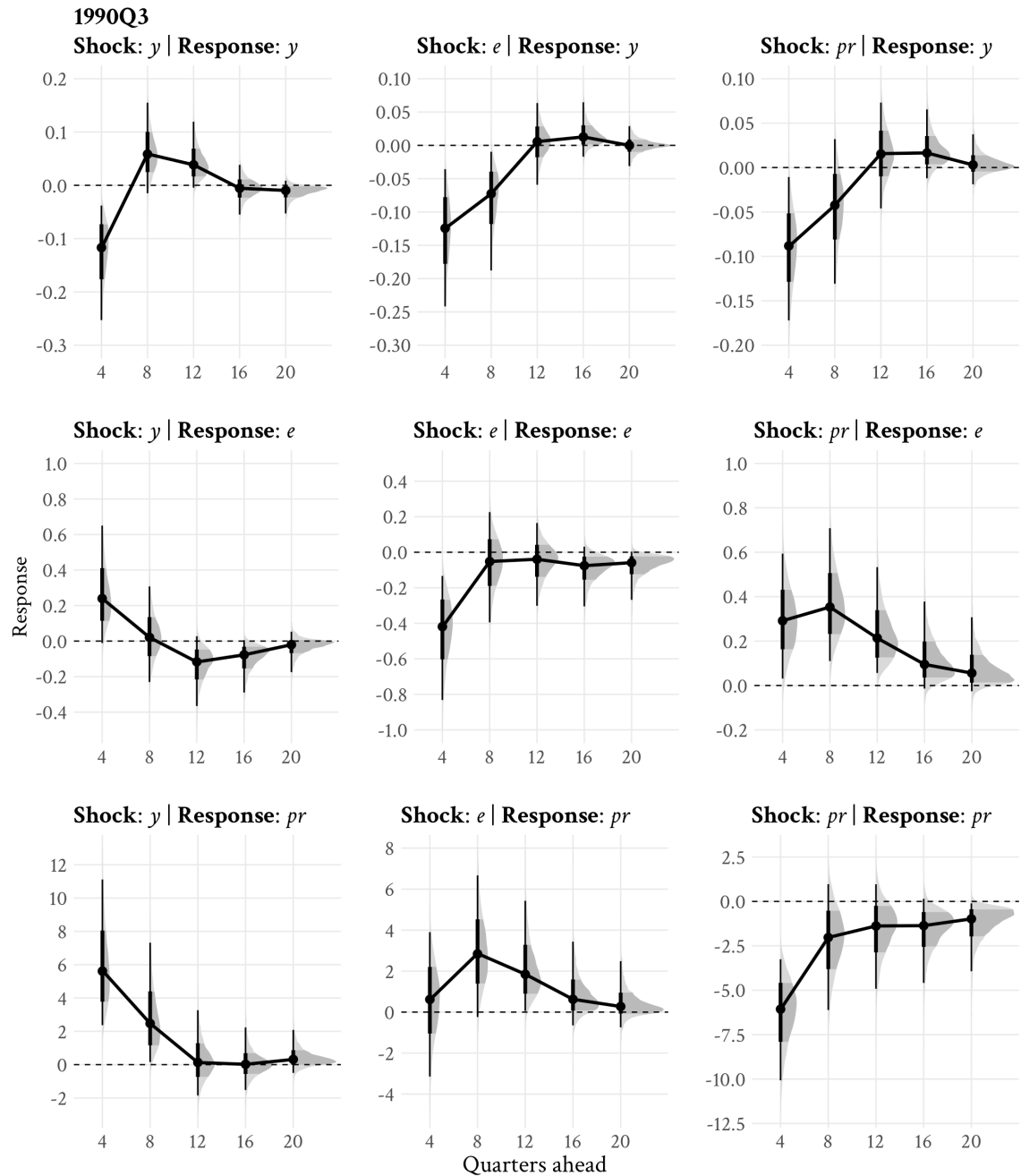

Figure 3: Posterior impulse-response functions in 1990Q3, baseline model. Note 1: All shocks denote a negative, one standard deviation exogenous innovation to the model variables. Note 2: Black dots denote posterior medians; darker shaded areas mark the 66% posterior density region; and the solid line connects posterior medians.

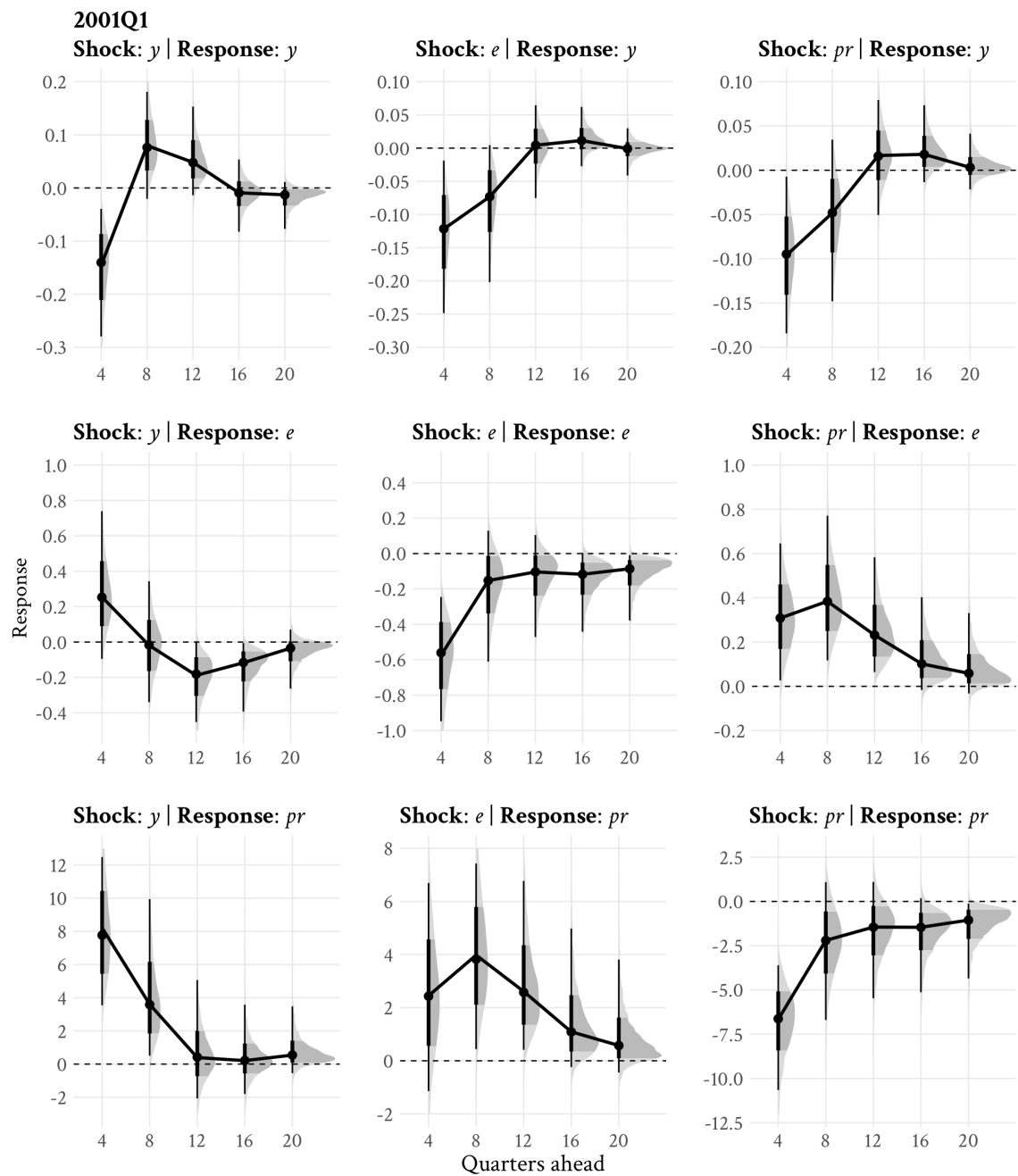

Figure 4: Posterior impulse-response functions in 2001Q1, baseline model. Note 1: All shocks denote a negative, one standard deviation exogenous innovation to the model variables. Note 2: Black dots denote posterior medians; darker shaded areas mark the 66% posterior density region; and the solid line connects posterior medians.

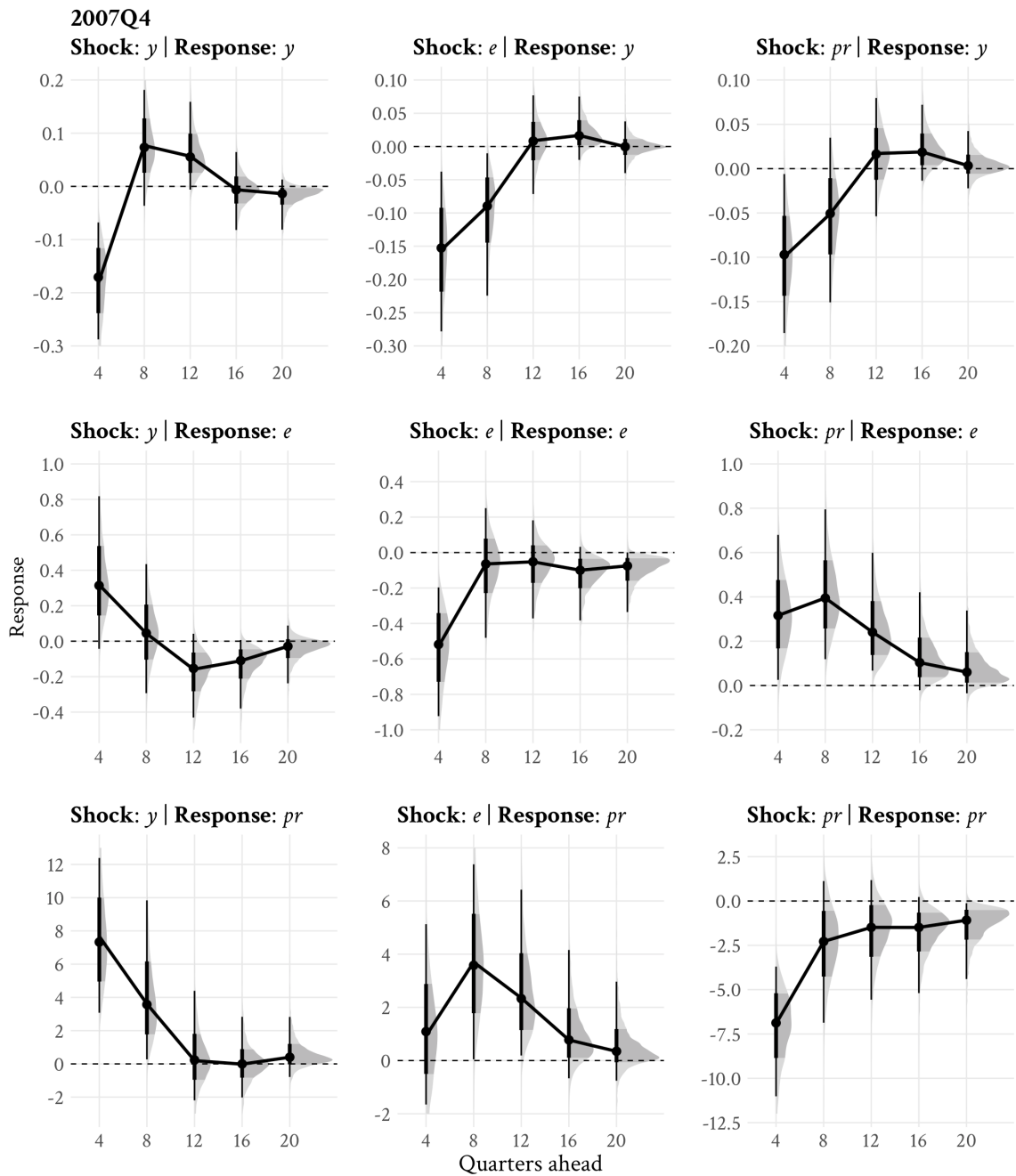

Figure 5: Posterior impulse–response functions in 2007Q4, baseline model. Note 1: All shocks denote a negative, one standard deviation exogenous innovation to the model variables. Note 2: Black dots denote posterior medians; darker shaded areas mark the 66% posterior density region; and the solid line connects posterior medians.

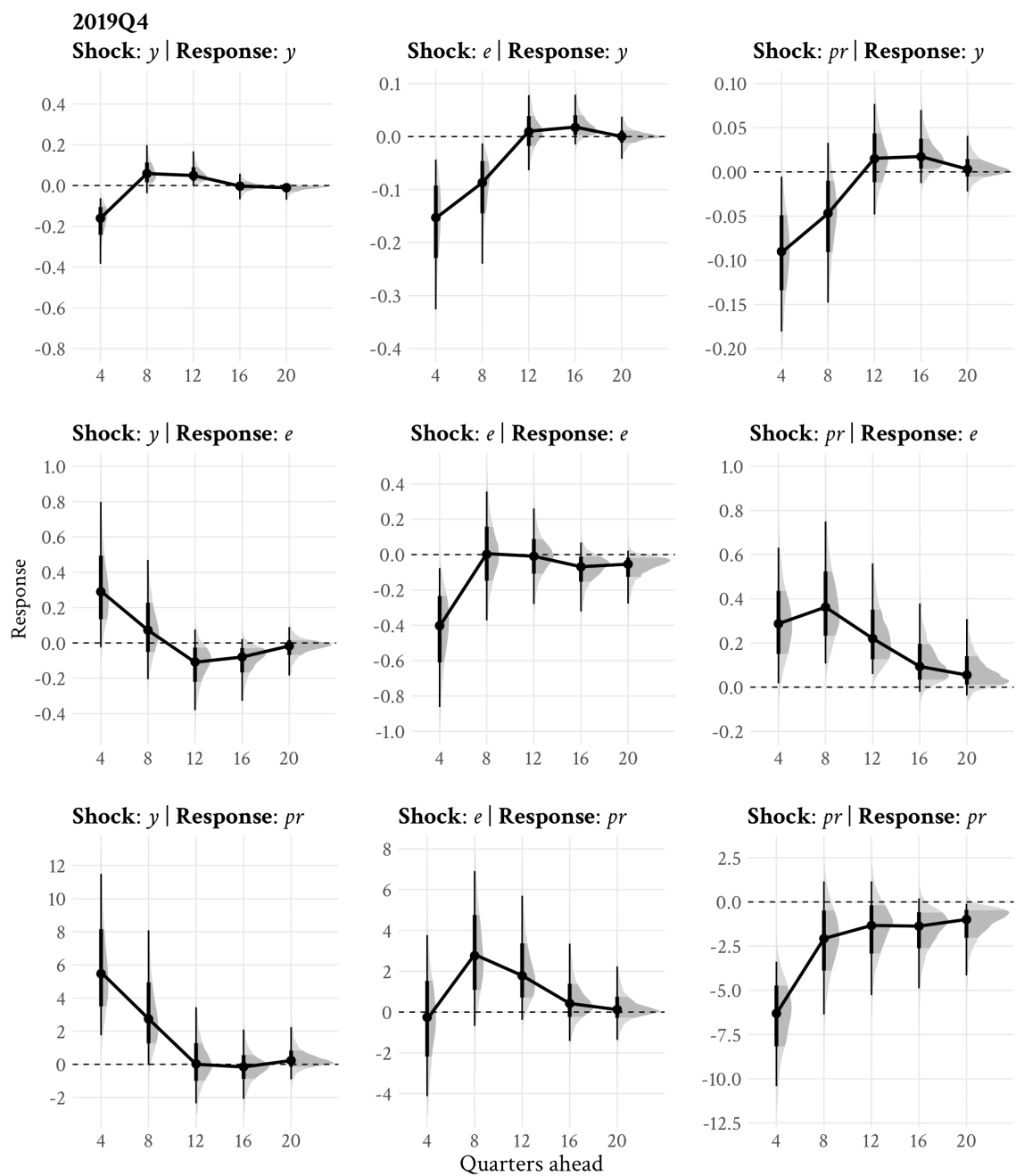

Figure 6: Posterior impulse-response functions in 2019Q4, baseline model. Note 1: All shocks denote a negative, one standard deviation exogenous innovation to the model variables. Note 2: Black dots denote posterior medians; darker shaded areas mark the 66% posterior density region; and the solid line connects posterior medians.

# B Baseline model's rebound charts for trough periods

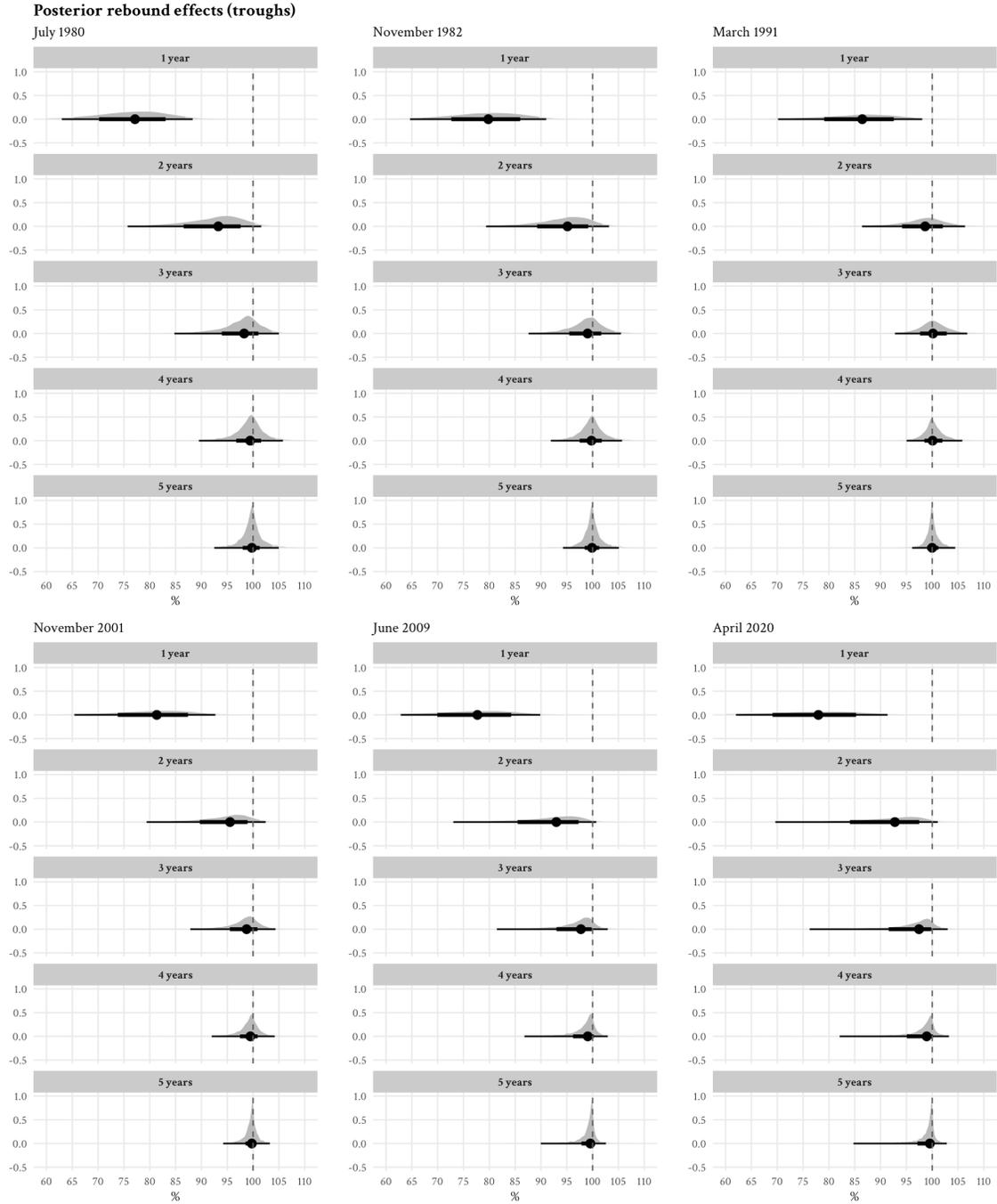

Figure 7: Posterior rebound effects at 1-, 2-, 3-, 4-, and 5-year intervals (business-cycle trough months). Note: Darker shaded areas denote the 90% density region.

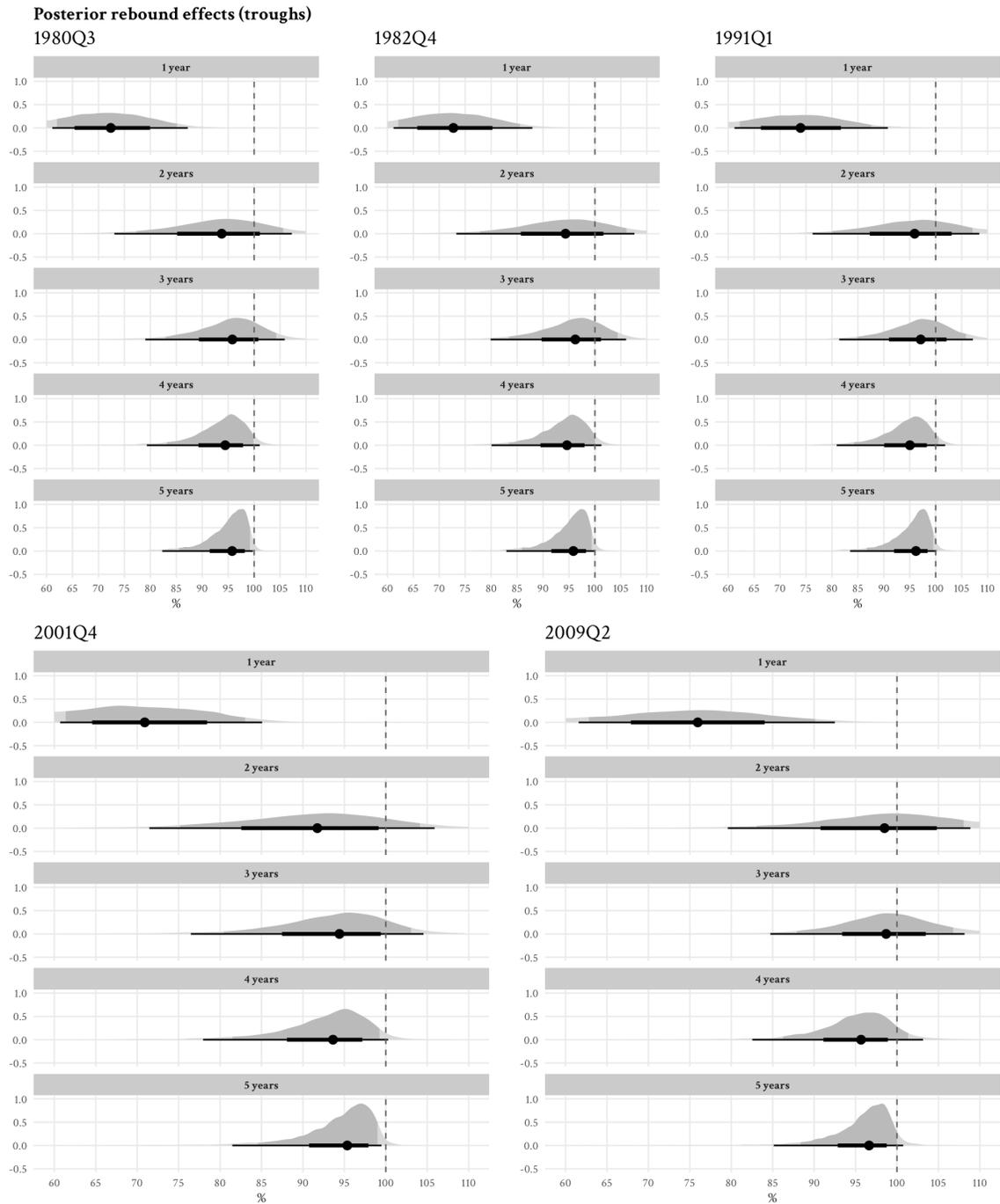

Figure 8: Posterior rebound effects at 1-, 2-, 3-, 4-, and 5-year intervals (business-cycle trough quarters). Note: Darker shaded areas denote the 90% density region.

# C Economy-wide rebound effect: Alternative specifications charts

## C.1 Model with $y^2$ (monthly data)

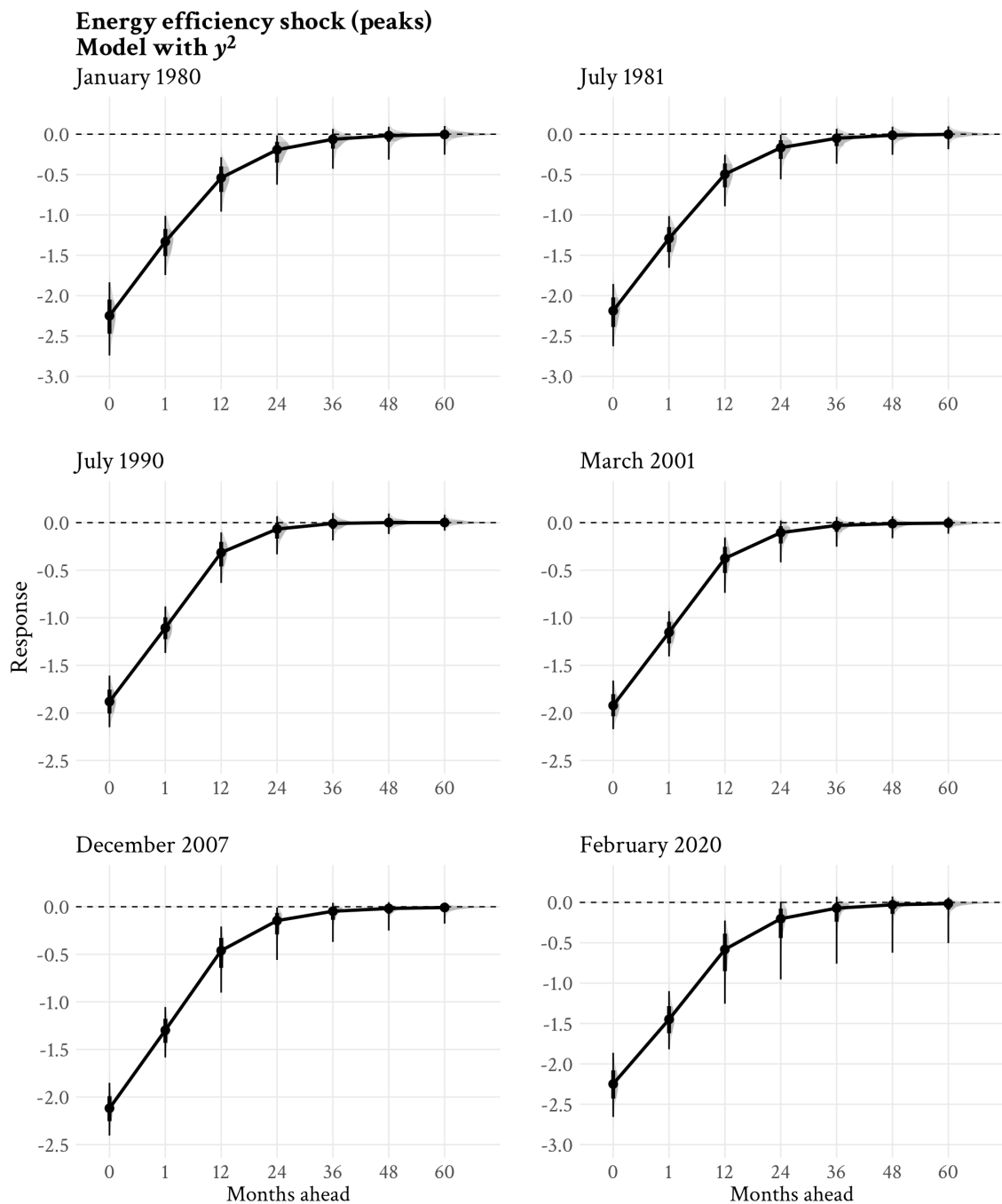

Figure 9: Posterior responses of energy use to a one-standard deviation shock in energy efficiency at different NBER-dated business-cycle peak months. Note: Black dots denote posterior medians; darker shaded areas mark the 66% posterior density region; and the solid line connects posterior medians.

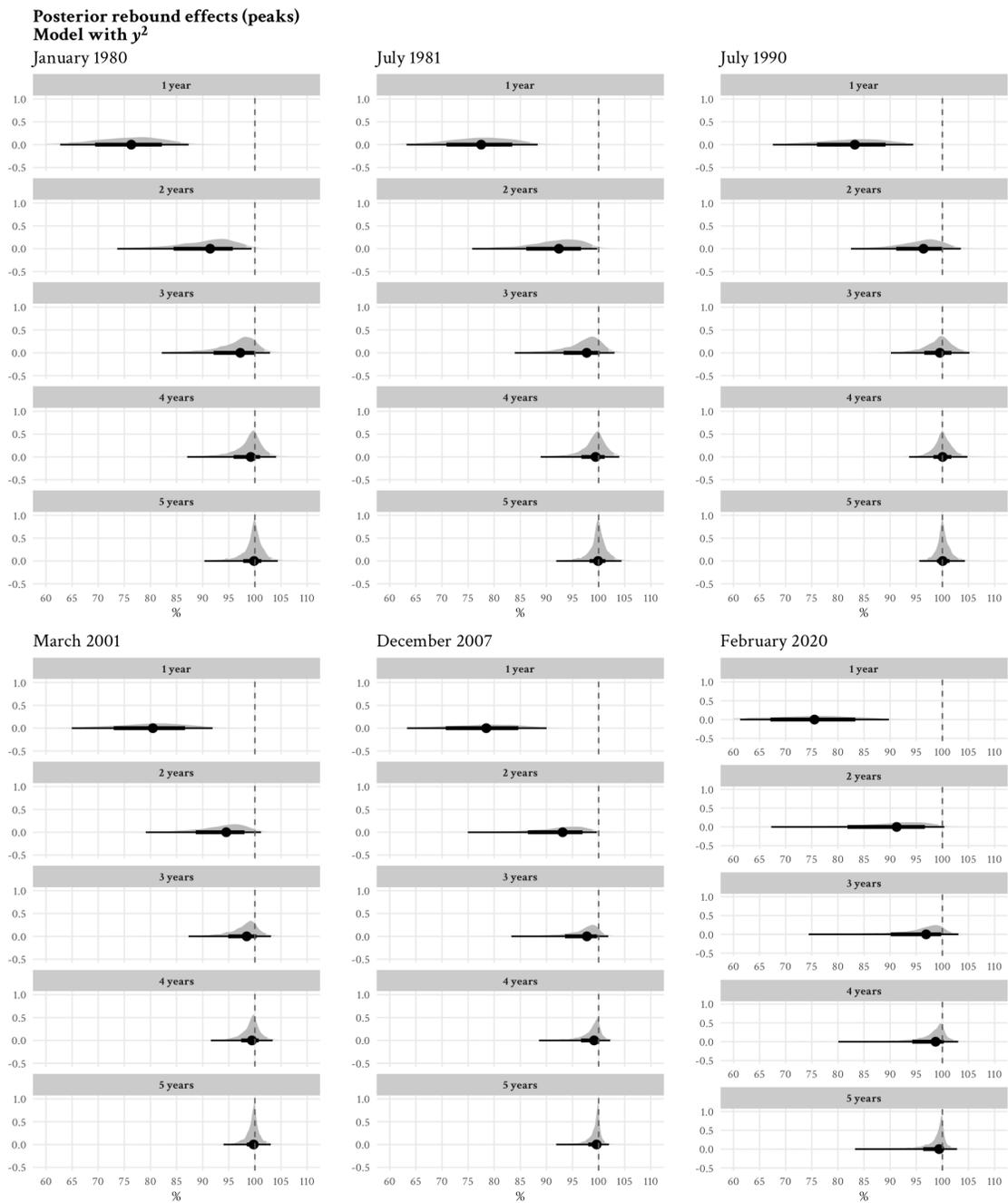

Figure 10: Posterior rebound effects at 1-, 2-, 3-, 4-, and 5-year intervals (business-cycle peak months). Note: Darker shaded areas denote the 90% density region.

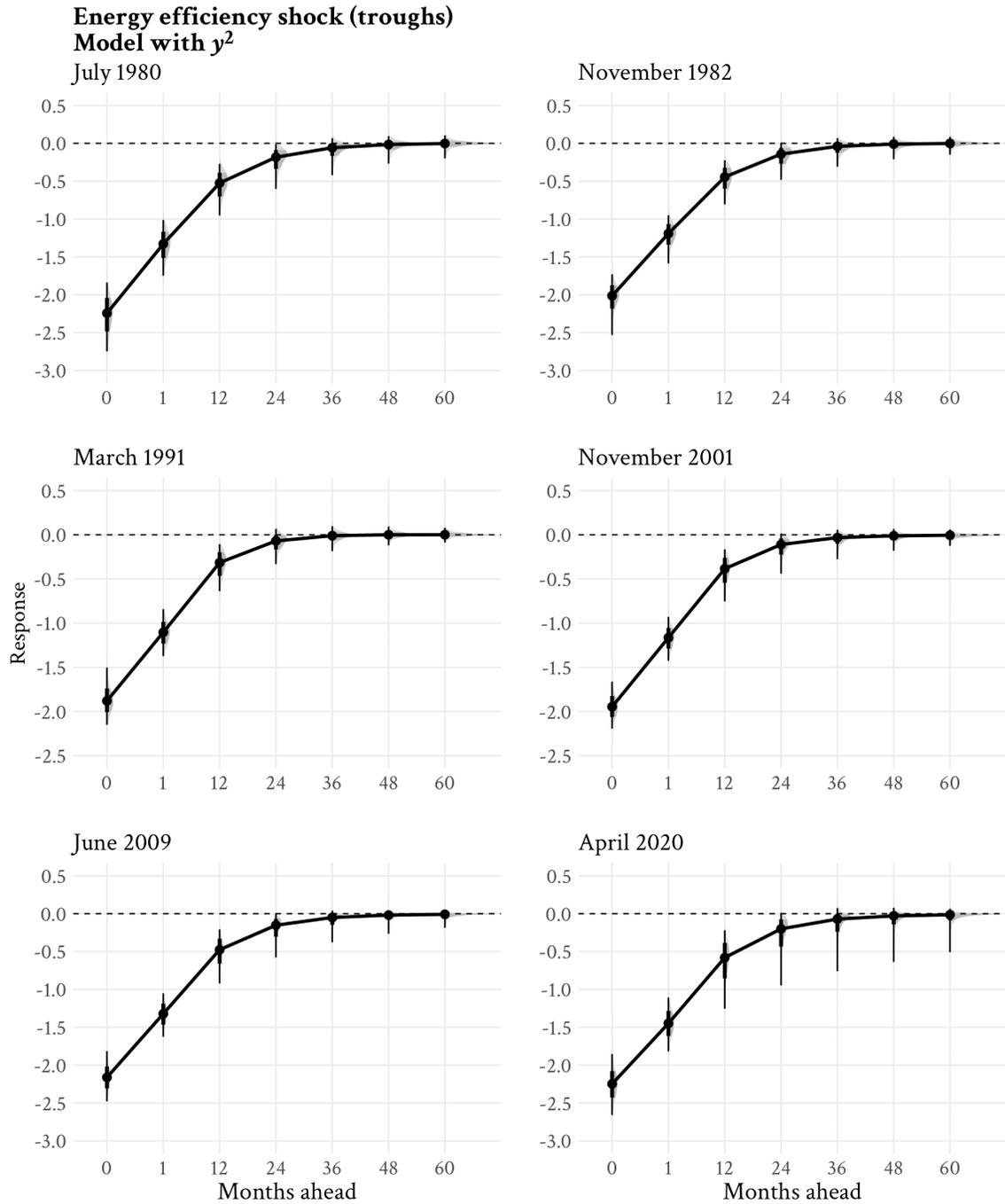

Figure 11: Posterior responses of energy use to a one-standard deviation shock in energy efficiency at different NBER-dated business-cycle trough months. Note: Black dots denote posterior medians; darker shaded areas mark the 66% posterior density region; and the solid line connects posterior medians.

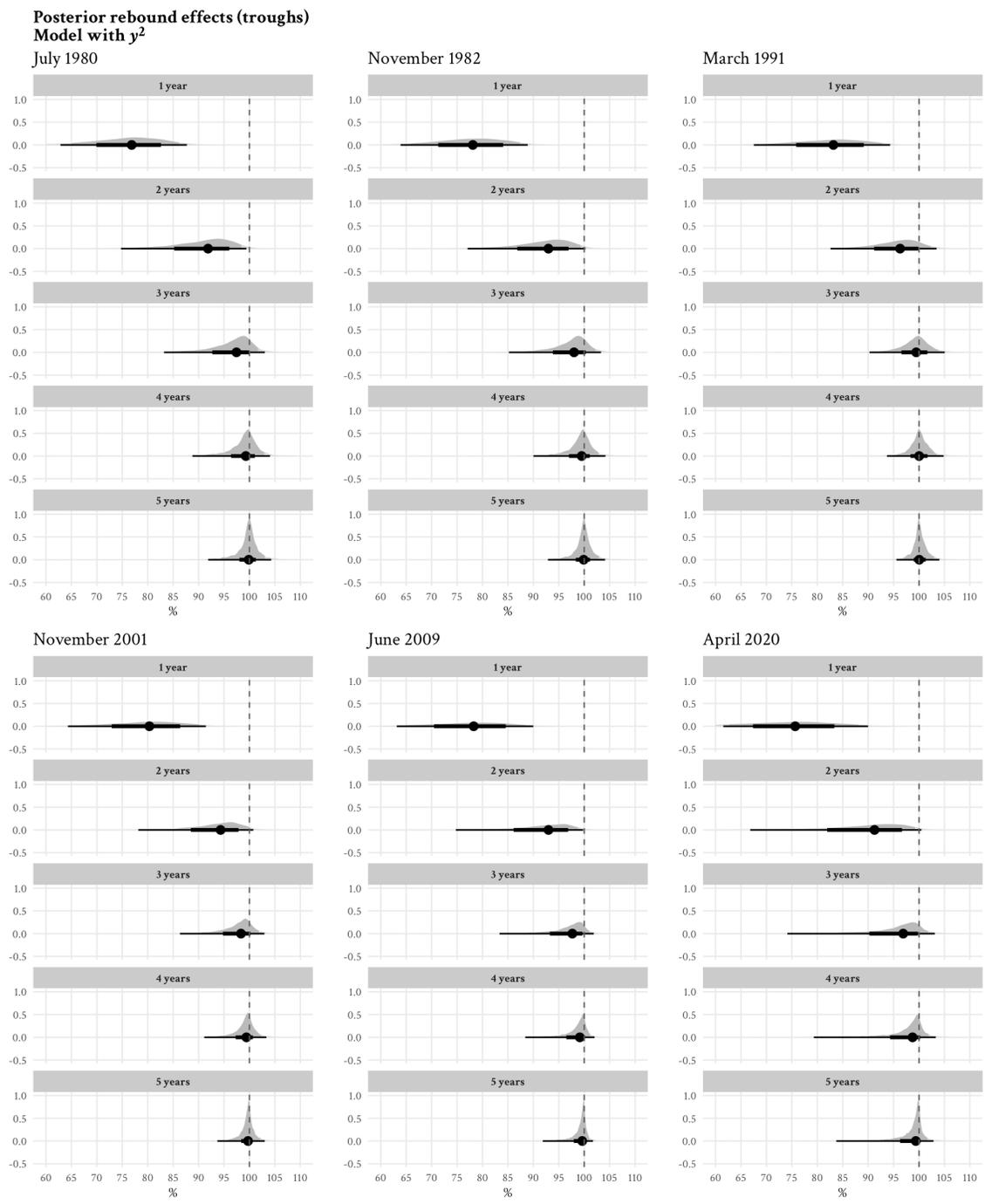

Figure 12: Posterior rebound effects at 1-, 2-, 3-, 4-, and 5-year intervals (business-cycle trough months). Note: Darker shaded areas denote the 90% density region.

## C.2 Model with $y^2$ (quarterly data)

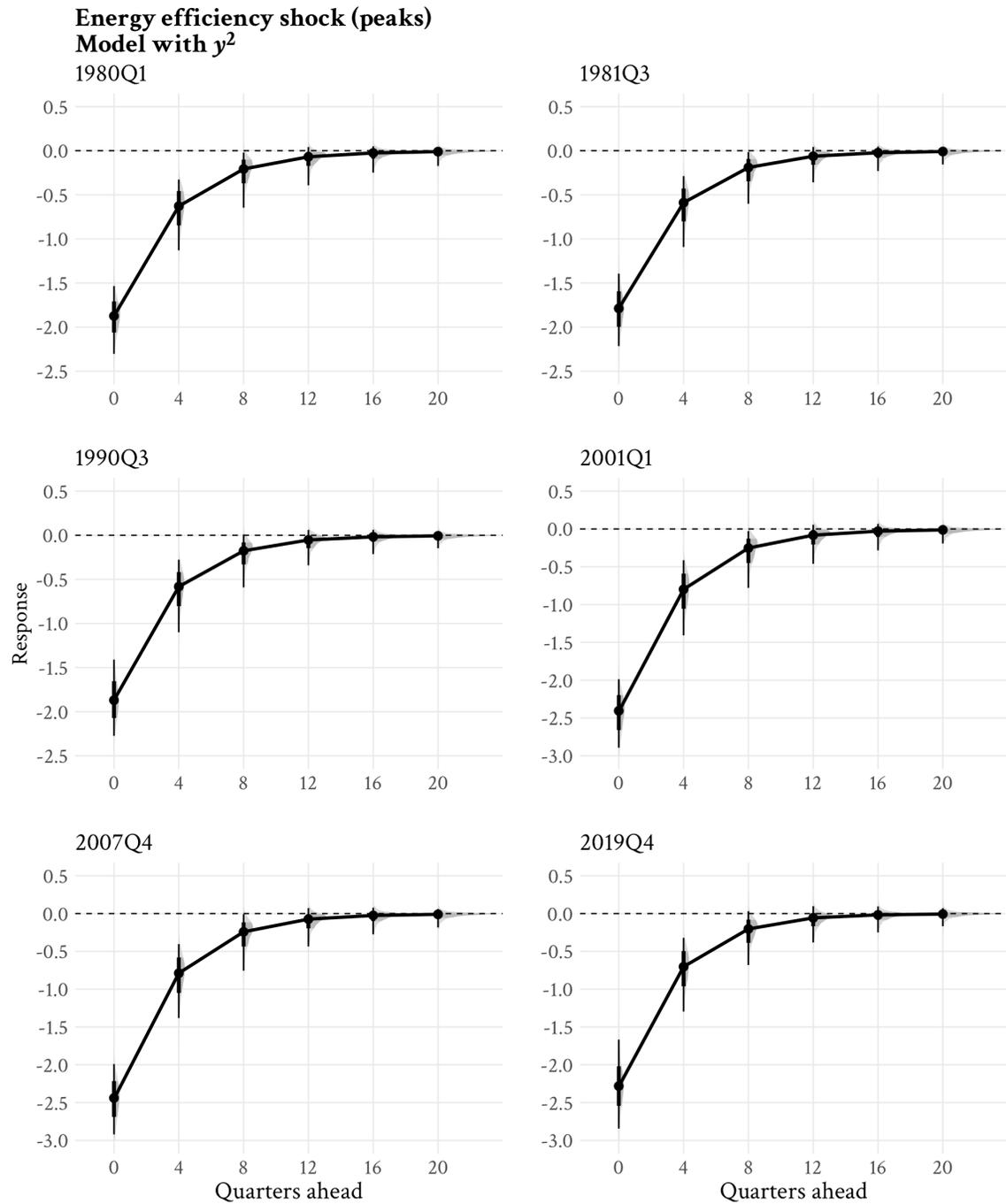

Figure 13: Posterior responses of energy use to a one-standard deviation shock in energy efficiency at different NBER-dated business-cycle peak quarters. Note: Black dots denote posterior medians; darker shaded areas mark the 66% posterior density region; and the solid line connects posterior medians.

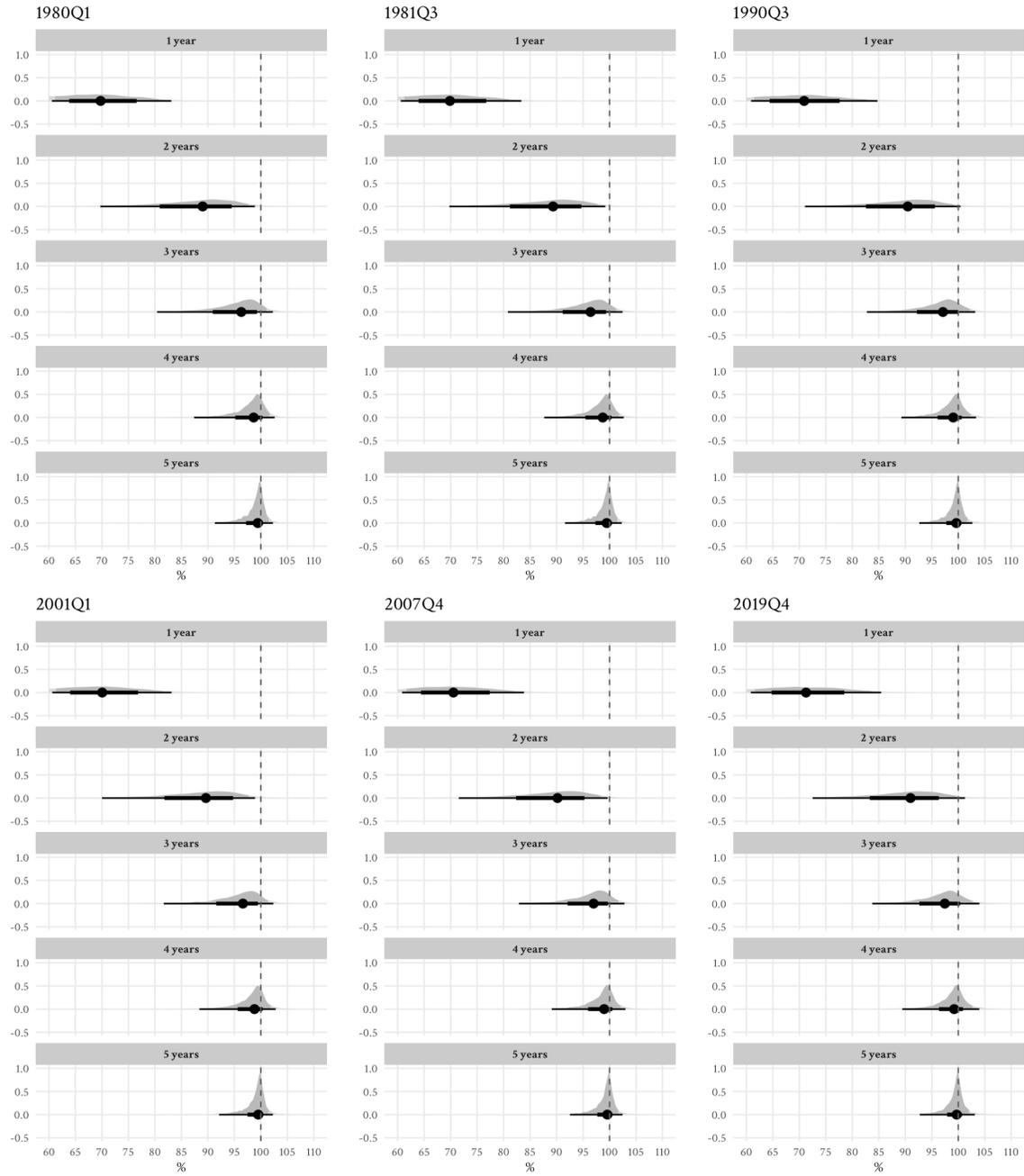

Figure 14: Posterior rebound effects at 1-, 2-, 3-, 4-, and 5-year intervals (business-cycle peak quarters). Note: Darker shaded areas denote the 90% density region.

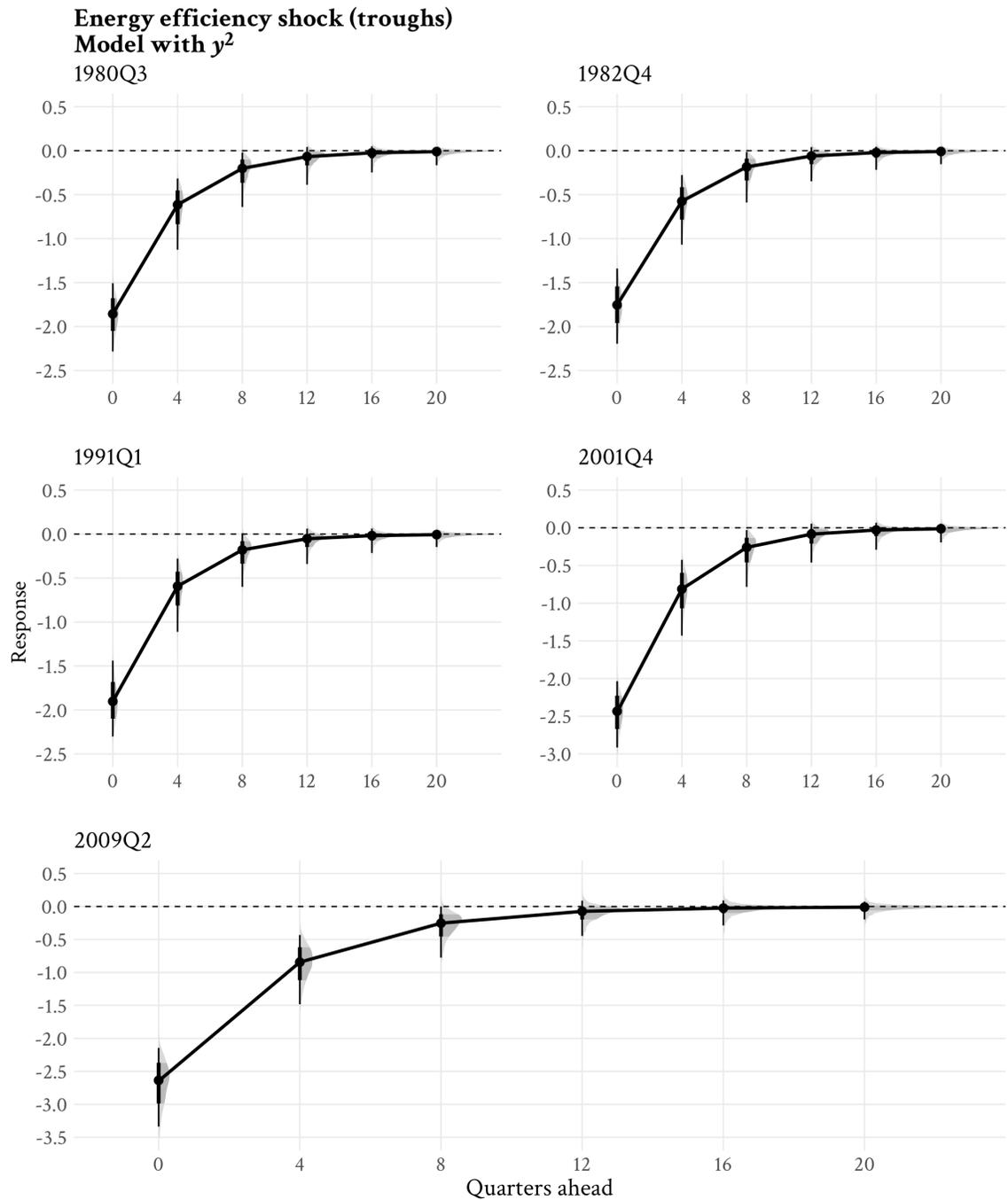

Figure 15: Posterior responses of energy use to a one-standard deviation shock in energy efficiency at different NBER-dated business-cycle trough quarters. Note: Black dots denote posterior medians; darker shaded areas mark the 66% posterior density region; and the solid line connects posterior medians.

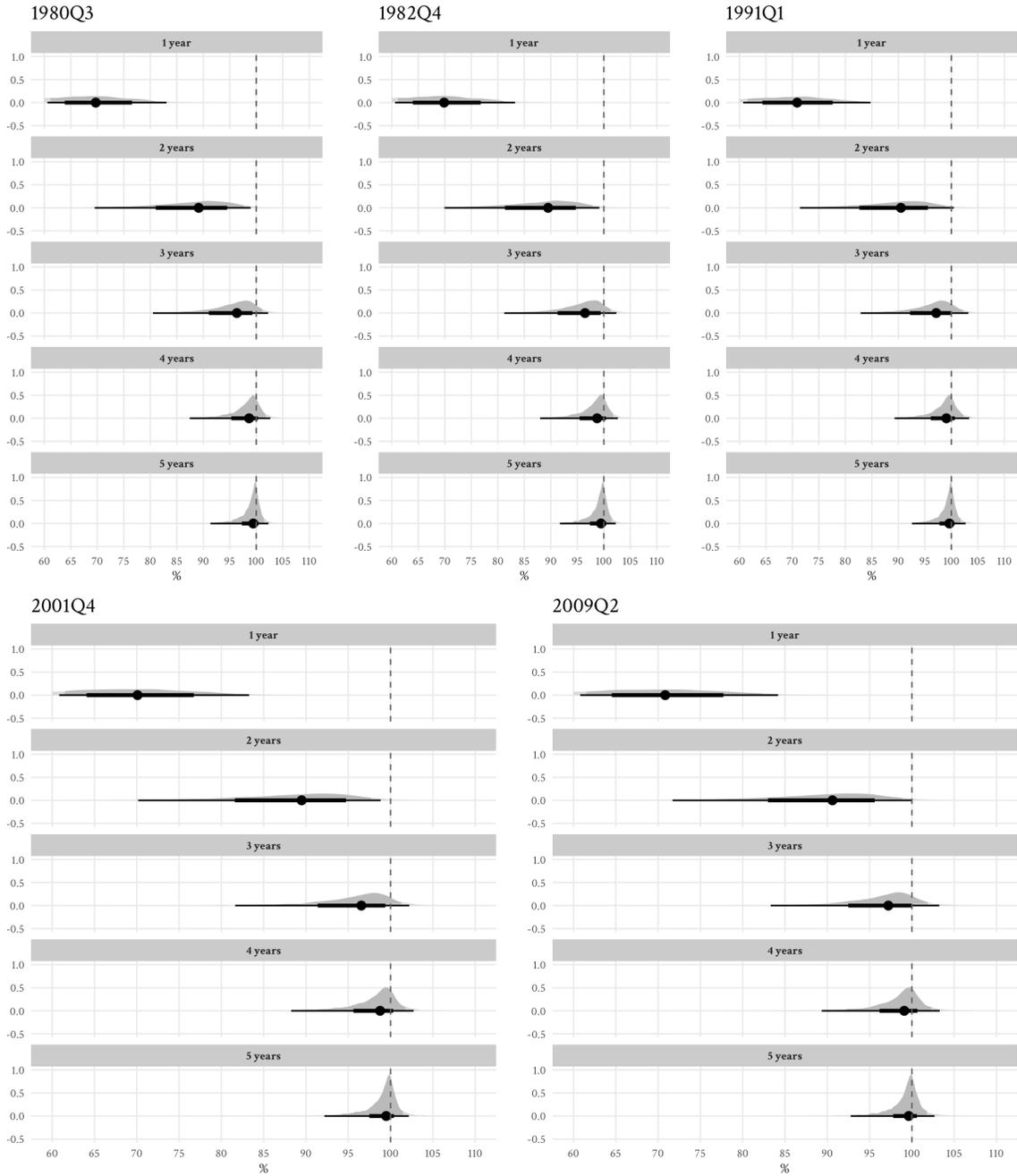

Figure 16: Posterior rebound effects at 1-, 2-, 3-, 4-, and 5-year intervals (business-cycle trough quarters). Note: Darker shaded areas denote the 90% density region.

## C.3 Model with $y^3$

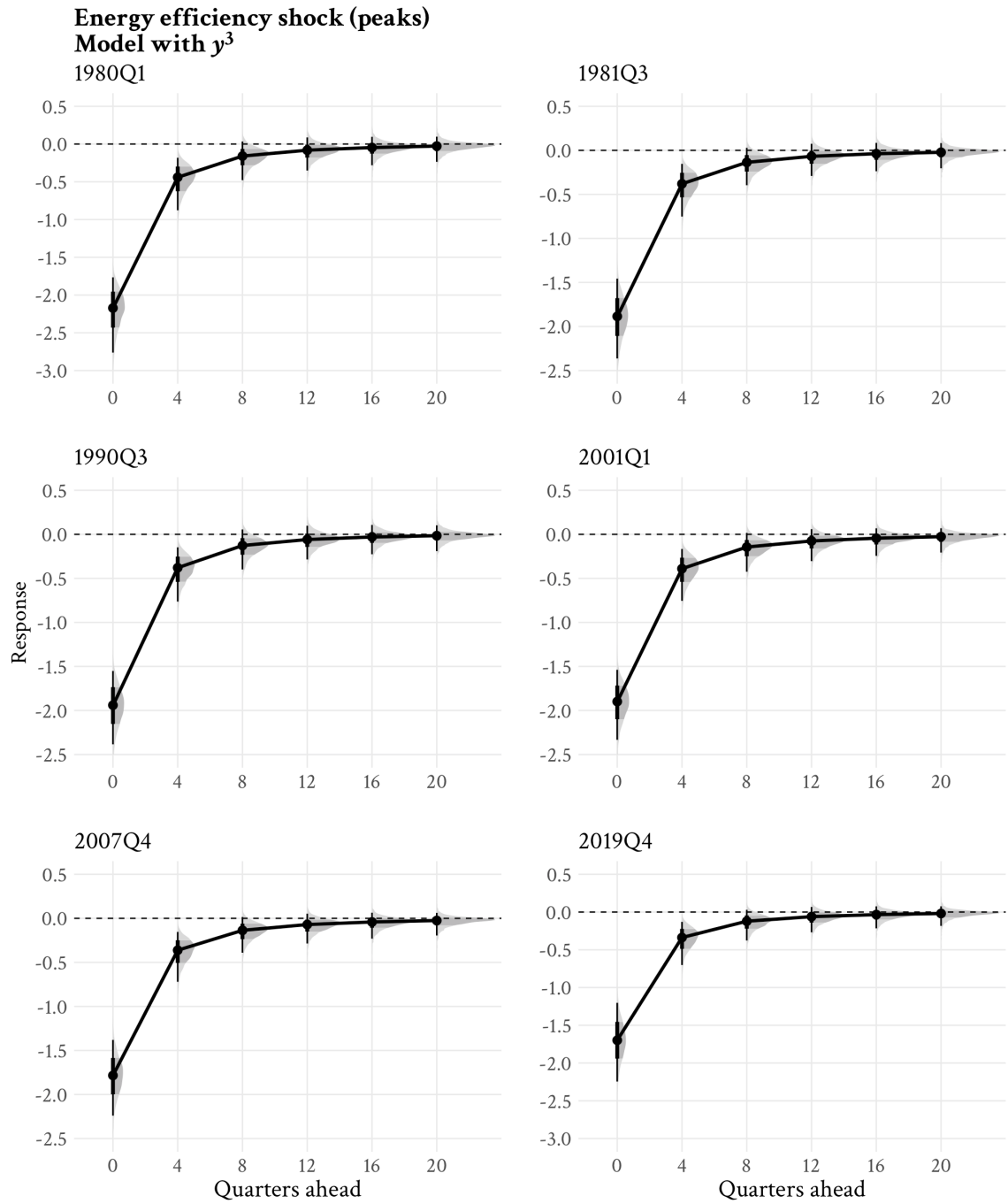

Figure 17: Posterior responses of energy use to a one-standard deviation shock in energy efficiency at different NBER-dated business-cycle peak quarters. Note: Black dots denote posterior medians; darker shaded areas mark the 66% posterior density region; and the solid line connects posterior medians.

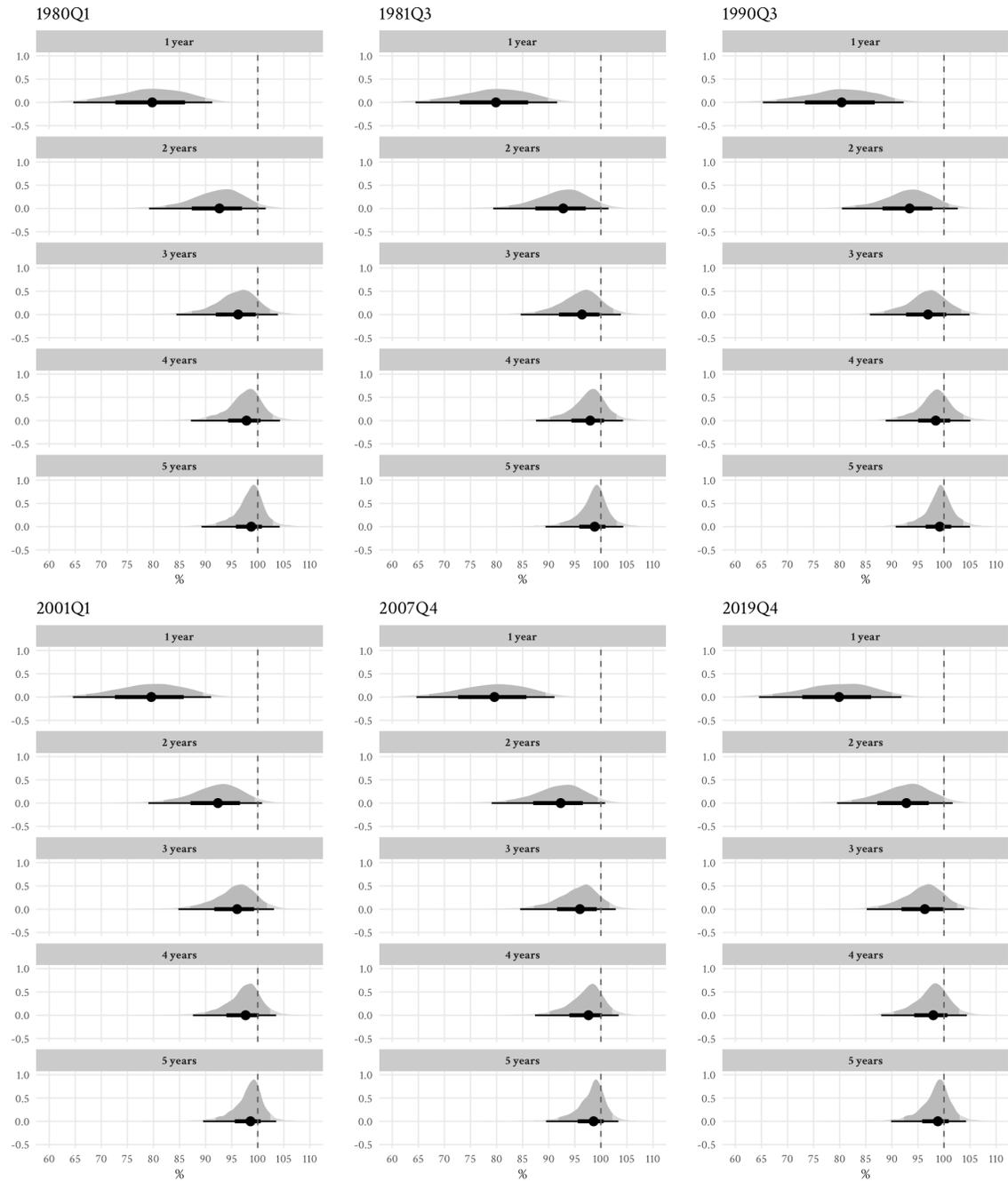

Figure 18: Posterior rebound effects at 1-, 2-, 3-, 4-, and 5-year intervals (business-cycle peak quarters). Note: Darker shaded areas denote the 90% density region.

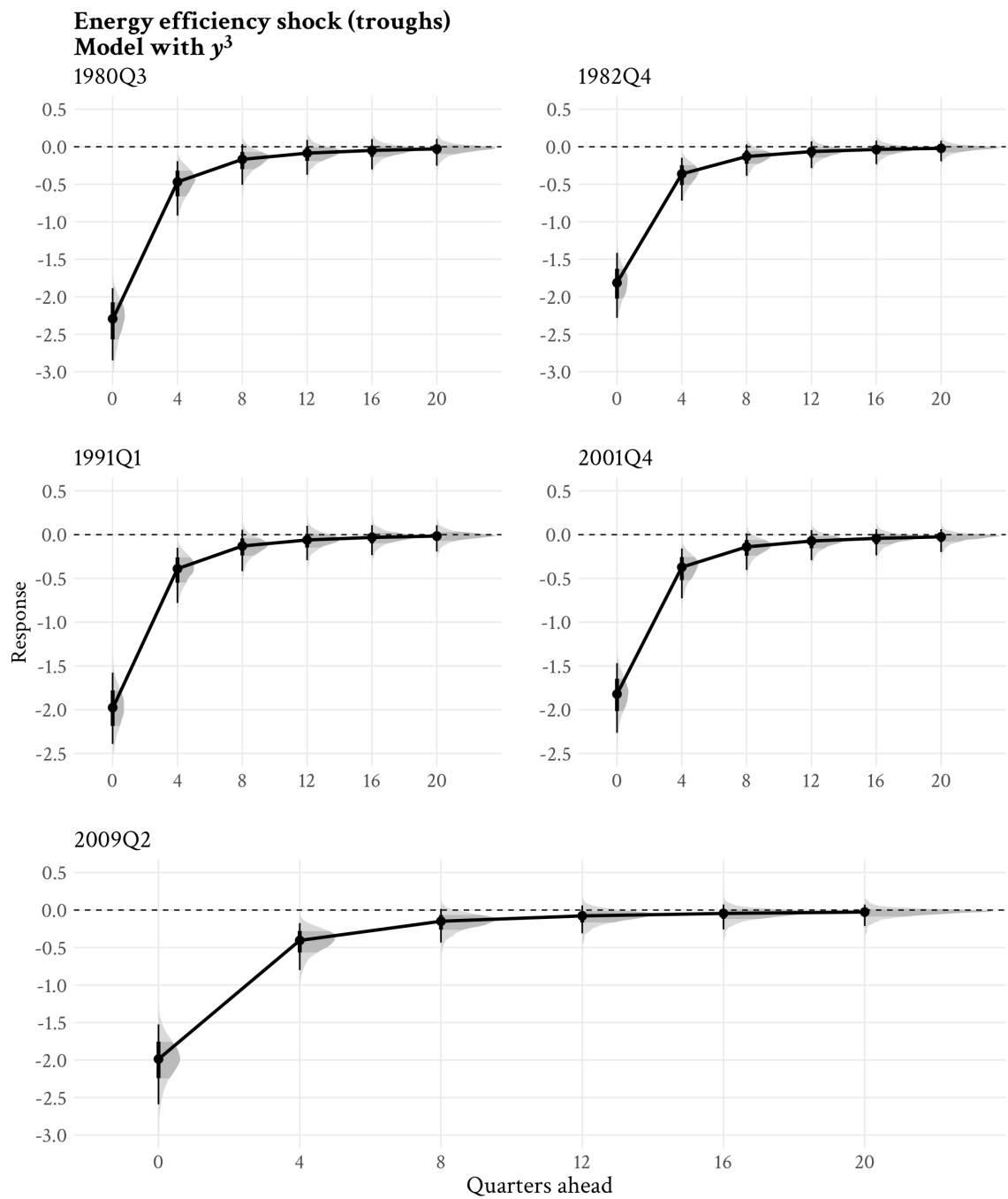

Figure 19: Posterior responses of energy use to a one-standard deviation shock in energy efficiency at different NBER-dated business-cycle trough quarters. Note: Black dots denote posterior medians; darker shaded areas mark the 66% posterior density region; and the solid line connects posterior medians.

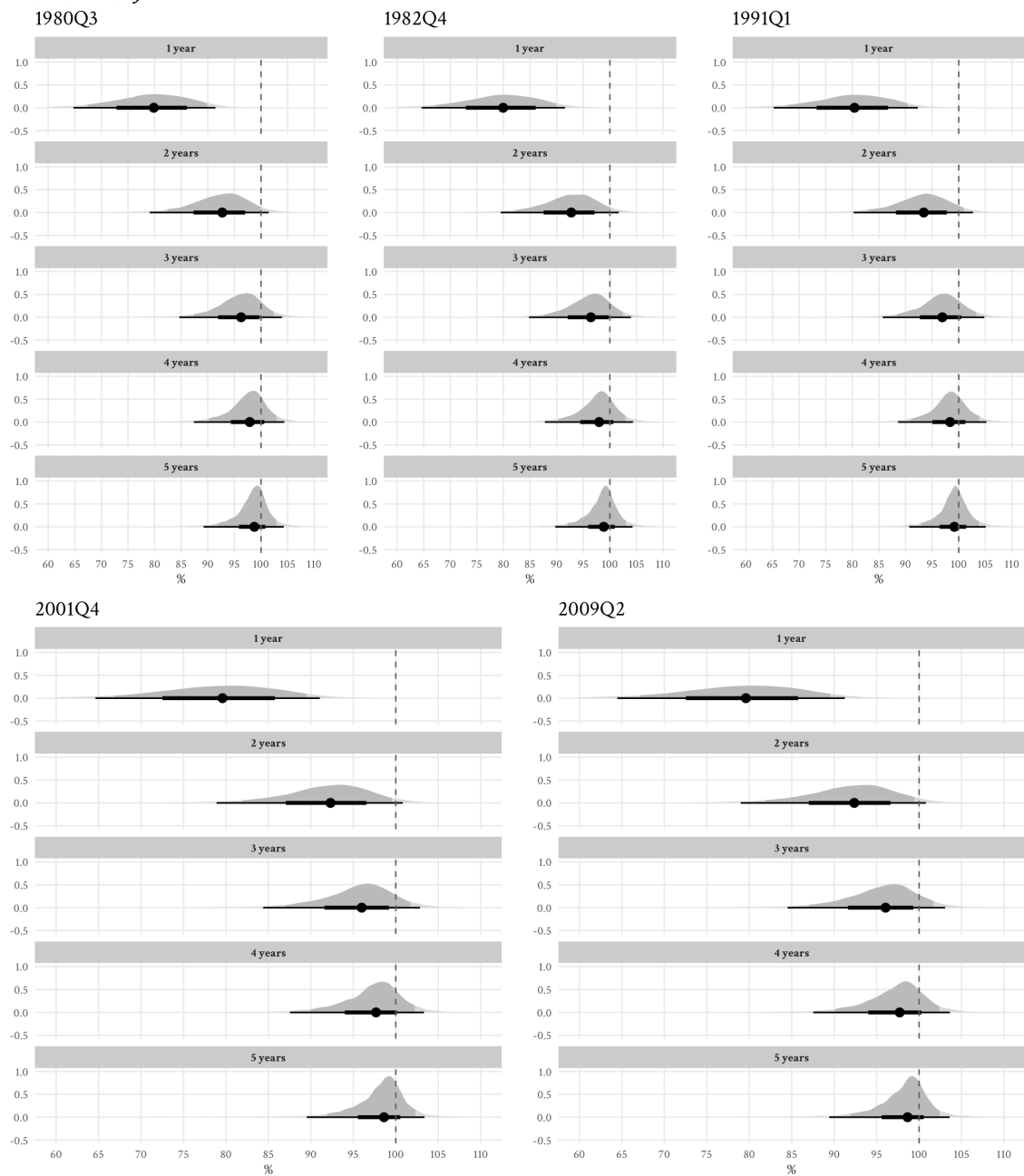

Figure 20: Posterior rebound effects at 1-, 2-, 3-, 4-, and 5-year intervals (business-cycle trough quarters). Note: Darker shaded areas denote the 90% density region.

# D Economy-wide rebound effect: Alternative specifications tables

## D.1 Model with $y^2$ (monthly data)

Table 1: Posterior median rebound effect, U.S. business-cycle trough months, model with $y^2$

| Horizon | 1980/07 | 1982/11 | 1991/03 | 2001/11 | 2009/06 | 2020/04 |
|---|---|---|---|---|---|---|
| 1 year  | 76.6 [66.4, 84.2]  | 78.0 [68.0, 85.7]  | 83.1 [73.0, 91.0]   | 80.2 [69.2, 88.3]  | 78.0 [66.1, 86.2]  | 74.1 [57.5, 85.2] |
| 2 years | 91.8 [81.9, 97.2]  | 92.9 [83.9, 98.1]  | 96.3 [88.9, 101.0]  | 94.3 [86.0, 98.9]  | 92.9 [82.6, 97.8]  | 91.0 [75.2, 97.9] |
| 3 years | 97.4 [90.0, 100.8] | 98.0 [91.6, 101.1] | 99.4 [95.0, 102.6]  | 98.3 [92.8, 100.9] | 97.7 [90.6, 100.2] | 96.8 [84.9, 100.5] |
| 4 years | 99.3 [94.3, 101.8] | 99.5 [95.3, 101.9] | 100.0 [97.3, 102.4] | 99.4 [95.9, 101.3] | 99.1 [94.5, 100.6] | 98.7 [90.3, 100.9] |
| 5 years | 99.9 [96.5, 102.1] | 99.9 [97.2, 101.8] | 100.0 [98.2, 102.0] | 99.7 [97.4, 101.2] | 99.6 [96.6, 100.5] | 99.3 [93.2, 100.8] |

*Note*: $10^{th}$ and $90^{th}$ percentile values in square brackets.

## D.2 Model with $y^2$ (quarterly data)

Table 2: Posterior median rebound effect, U.S. business-cycle trough quarters, model with $y^2$

| Horizon | 1980Q3 | 1982Q4 | 1991Q1 | 2001Q4 | 2009Q2 |
|---|---|---|---|---|---|
| 1 year  | 66.6 [52.4, 77.5]  | 67.1 [53.0, 77.8]        | 68.4 [54.4, 78.8]  | 67.0 [53.3, 77.8]  | 68.4 [54.9, 79.1] |
| 2 years | 89.1 [77.3, 95.9]  | 89.4 [77.7, 96.1]        | 90.5 [79.3, 97.1]  | 89.4 [77.7, 96.1]  | 90.6 [79.4, 97.2] |
| 3 years | 96.3 [88.4, 100.1] | 96.4 [88.7, 100.3]       | 97.1 [89.8, 100.9] | 96.5 [88.7, 100.1] | 97.2 [90.1, 100.8] |
| 4 years | 98.6 [93.5, 100.9] | 98.7 [93.7, 101.0]       | 99.1 [94.3, 101.4] | 98.8 [93.7, 100.9] | 99.1 [94.6, 101.3] |
| 5 years | 99.4 [95.9, 100.8] | 9death aro9.5 [96.0, 100.9] | 99.6 [96.5, 101.1] | 99.5 [96.0, 100.8] | 99.6 [96.6, 101.0] |

*Note*: $10^{th}$ and $90^{th}$ percentile values in square brackets.

## D.3 Model with $y^3$

Table 3: Posterior median rebound effect, U.S. business-cycle trough quarters, model with $y^3$

| Horizon | 1980Q3 | 1982Q4 | 1991Q1 | 2001Q4 | 2009Q2 |
|---|---|---|---|---|---|
| 1 year  | 79.7 [69.5, 87.9]  | 79.8 [69.6, 87.9]  | 80.3 [70.3, 88.5]  | 79.4 [69.4, 87.5]  | 79.5 [69.3, 87.5] |
| 2 years | 92.7 [85.2, 98.4]  | 92.7 [85.3, 98.4]  | 93.4 [85.9, 99.3]  | 92.3 [84.7, 97.9]  | 92.4 [84.7, 98.1] |
| 3 years | 96.3 [90.0, 100.9] | 96.4 [90.1, 101.0] | 96.9 [90.8, 101.8] | 96.0 [89.6, 100.3] | 96.1 [89.7, 100.5] |
| 4 years | 97.9 [92.5, 101.7] | 98.0 [92.7, 101.8] | 98.4 [93.3, 102.5] | 97.7 [92.3, 101.1] | 97.7 [92.3, 101.2] |
| 5 years | 98.8 [94.2, 101.8] | 98.9 [94.4, 101.8] | 99.2 [94.9, 102.4] | 98.6 [94.0, 101.3] | 98.6 [94.0, 101.3] |

*Note*: $10^{th}$ and $90^{th}$ percentile values in square brackets.